\documentclass[draft]{agujournal2019}
\usepackage{url} 
\usepackage{lineno}
\usepackage[inline]{trackchanges} 

\usepackage{soul}

\draftfalse

\journalname{Journal of Advances in Modeling Earth Systems (JAMES)}

\begin{document}

\justify

\title{Non-local parameterization of atmospheric subgrid processes with neural networks}

%
%




\authors{Peidong Wang\affil{1}, Janni Yuval\affil{1}, and Paul A. O'Gorman\affil{1}}
\affiliation{1}{Department of Earth, Atmospheric and Planetary Sciences, Massachusetts Institute of Technology, Cambridge, Massachusetts 02139, USA}

\correspondingauthor{Peidong Wang}{pdwang@mit.edu}
\correspondingauthor{Janni Yuval}{janniy@mit.edu}

\begin{keypoints}
\item Using non-local inputs from neighboring atmospheric columns improves offline performance of a neural network parameterization
\item Improvements in mid-latitudes are associated with cases with mid-latitude fronts, where subgrid squall-line features are present
\item An explainable artificial intelligence technique shows non-local winds are especially useful for parameterizing subgrid momentum transport
\end{keypoints}

%
%

%
%


 \begin{abstract}

 Subgrid processes in global climate models are represented by parameterizations which are a major source of uncertainties in simulations of climate. In recent years, it has been suggested that machine-learning (ML) parameterizations based on high-resolution model output data could be superior to traditional parameterizations.
 Currently, both traditional and ML parameterizations of subgrid processes in the atmosphere are based on a single-column approach, which only use information from single atmospheric columns.
However, single-column parameterizations might not be ideal since certain atmospheric phenomena, such as organized convective systems, can cross multiple grid boxes and involve slantwise circulations that are not purely vertical.
Here we train neural networks (NNs) using non-local inputs spanning over 3$\times$3
columns of inputs. We find that including the non-local inputs improves the offline
prediction of a range of subgrid processes.
The improvement is especially notable for subgrid momentum transport and for atmospheric conditions associated with mid-latitude fronts and convective instability.
Using an interpretability method, we find that the NN improvements partly rely on using the horizontal wind divergence, and we further show that including the divergence or vertical velocity as a separate input substantially improves offline performance. 
However, non-local winds continue to be useful inputs for parameterizating subgrid momentum transport even when the vertical velocity is included as an input.
Overall, our results imply that the use of non-local variables and the vertical velocity as inputs  could improve the performance of ML parameterizations, and the use of these inputs should be tested in online simulations in future work.
 \end{abstract}

 \section*{Plain Language Summary}

Current global climate models cannot resolve small-scale processes, such as clouds and convection, which are crucial for accurate simulations of climate, and the effect of these processes is approximated using parameterizations. Traditionally, these parameterizations rely on simple conceptual models, but in recent years machine learning has also been used to develop new parameterizations.
Both traditional and machine learning parameterizations rely on a simple approach in which the vertical structure of a single atmospheric column is used to predict the effect of unresolved small-scale processes on the column itself. 
Here we use machine learning to show that this single-column approach might hamper the accuracy of parameterizations.
We demonstrate that a machine-learning parameterization that uses information from multiple atmospheric columns simultaneously (rather than information from a single atmospheric column) better predicts the effects of small-scale processes compared to the same approach but only using information from a single column.
We show that non-local inputs are especially important for parameterizing subgrid momentum transport and for mid-latitude situations with atmospheric fronts.
Including only neighboring columns is sufficient to improve the parameterization in climate model simulations, and therefore the increase in computational expense should not be a barrier in the implementation of non-local parameterizations.

%
%
 \section{Introduction}
 \label{intro}
 
 Accurate climate projections are of great societal relevance (e.g. in assessing the risk from heavy rainfall events) and scientific interest (e.g. in understanding the dynamics of the climate system). These projections rely on global climate models that typically have grid spacing of a few tens to a hundred kilometers and thus, cannot resolve processes that occur on smaller scales (i.e., subgrid processes). Because subgrid processes, such as convection and clouds, have important consequences for Earth's climate, there is a need to represent them using parameterizations. These parameterizations approximate the effects of unresolved processes on the resolved fields. Traditional parameterizations rely partly on physics but also on simple conceptual models and heuristic approximations, and they are a major source of  uncertainties in climate models and climate projections \cite{Wilcox_2007_extremeprecp, Sherwood_2014_climatesensitivity, Bony_2015, Tapio_2017}. 
 
 In recent years, it has been suggested that data-driven parameterizations based on machine learning (ML) could be computationally more efficient and/or more accurate than traditional parameterizations \cite{krasnopolsky2013using,gentine2018could,Ogorman2018using}. 
For example, several studies have used output from the super-parameterized Community Atmosphere Model (SPCAM) to emulate its super-parameterization for aquaplanets and more realistic configurations \cite{Rasp_2018,Han_2020_GCM,mooers2020assessing}. Other studies have learned from output from three-dimensional high-resolution simulations which resolve processes that are usually subgrid in global climate models 
\cite{brenowitz2018prognostic,Brenowitz_2019,Yani_natcommun,Yani_2021GRL}. In this approach, the output from the high-resolution simulation is first coarse-grained (i.e., averaged onto a coarser grid), and then an ML algorithm is used to predict the effect of the small-scale processes on the (coarse-grained) prognostic variables.

Both traditional and ML-based parameterizations of subgrid atmospheric processes rely on a single column framework. In this framework, the vertical profiles of the moisture, temperature and winds 
in an atmospheric column are typically used as the inputs to the parameterization. In turn, the parameterization predicts the effect of small-scale processes on the resolved prognostic fields in the atmospheric column.
The motivation to use single-column parameterization relies on the idea that the subgrid processes primarily rearrange mass, momentum and energy in the vertical, and that a single-column framework is adequate to model these vertical processes \cite{STENSRUD_local_param}. In addition, a single column formulation does not require horizontal communication between columns which may be stored on different processors, thus increasing computational efficiency and simplifying code development. Thus, traditional and ML parameterizations for the atmosphere have been based on a single-column approach, with the exception of some stochastic parameterizations in which the stochasticity is implemented non-locally \cite{Palmer_nonlocal}. We note also that an ML parameterization has been developed that includes non-locality in time \cite{Han_2020_GCM} which may be related to non-locality in space for propagating weather systems.

However, a single-column parameterization structure may not be ideal when predicting the effect of certain subgrid atmospheric processes or for certain atmospheric conditions. For example, slantwise convection, which is related to conditional symmetric instability that is prevalent in the subtropics and mid-latitudes \cite{Chen_MPV}, is not a purely vertical process and may not be well predicted using only inputs from a single vertical column.   
Moreover, mesoscale convective systems are driven by small-scale convective processes which are not resolved by current global climate models, but are organized in coherent tilted structures that are larger than a single grid box of a global climate model
\cite{Houze_mesoscale_convection}. 
In such multiscale convective systems,  unresolved convective processes in an atmospheric column might have some statistical relation with the atmospheric state of neighboring columns, and therefore including  information from neighboring columns in parameterizations could potentially improve their performance. In addition, knowledge about the three dimensional structure of the winds may help in estimating the magnitude and direction of subgrid momentum transport.

The examples mentioned above provide the motivation for this study in which we test whether using non-local information in the horizontal could improve ML parameterizations and potentially also traditional parameterizations. We learn from coarse-grained output of a high-resolution three-dimensional simulation, noting that the alternative of emulating a super-parameterization is less attractive in this case since the super-parameterization already imposes a single-column structure. 
In related recent work, a non-local convolutional neural network (NN) was used to predict the horizontal subgrid eddy momentum forcing for an ocean gyre circulation using inputs over 40$\times$40 horizontal blocks of grid boxes \cite{Zanna_2019}. Here, given our focus on convection, we use inputs from the 3$\times$3 grid columns in the immediate neighborhood of the target column.  In a distributed computing environment, such neighboring inputs would typically be available on a given processor through a halo around the subdomain for that processor, and thus using only 3$\times$3 columns helps to limit inter-processor communication and computational expense.

We also investigate which non-local features are important to improve the predictions of the parameterization.
We find that the non-local variables that ML parameterizations rely on are mostly the wind fields. For some outputs, the patterns of inputs used suggest that the parameterization is learning the horizontal wind divergence, which is related to the vertical velocity through the mass continuity equation. This motivates us to also study the use of the (local) vertical velocity as an input to a single-column parameterization.
Some conventional convection schemes rely on closures in terms of the vertical velocity \cite{ooyama1969numerical} or the moisture convergence which is closely related to the vertical velocity \cite{Kuo_jars_1974}, although this is less common in climate models compared to closures based on measures of instability (see Table 2 of \citeA{pathak_precipitation_2019}).
Whether variables such as the vertical velocity or moisture convergence 
should be included as inputs to convection and cloud parameterizations is the subject of debate \cite{emanuel1994large,george2021}. 
In particular, using convergence as an input may lead to reverse causation 
since convergence is both a cause and a consequence of convection \cite{back2009}, and this is potentially an issue in the context of ML parameterizations which can be unstable as a result of learning non-causal relations between inputs and outputs \cite{Brenowitz_2020}.
Given the uncertainty as to whether the divergence or vertical velocity should be included as inputs in ML parameterizations, we choose to present results for parameterizations with and without these inputs.


 We organize the paper as follows. In section \ref{data_methods}, we describe the high-resolution simulation used to build our training and testing data sets (Section \ref{data_methods_simulation}), the NN parameterizations (Section \ref{data_methods_NN}), and an explainable ML technique, called layer-wise relevance propagation, that we use to interpret NN parameterizations (Section \ref{data_methods_LRP}). Then in Section \ref{results}, we begin to analyze the results by quantifying the performance of 
the parameterizations and comparing parameterizations that use non-local inputs to parameterizations that use a single column structure, with and without vertical wind as an input (Section \ref{results_identify_improv}). 
Next, we focus on mid-latitudes where non-local inputs are especially useful and identify atmospheric conditions in which the non-local parameterization substantially improves or does not improve the predictions (Section \ref{results_explain_improv}).  We use layer-wise relevance propagation to understand on which non-local inputs the NN parameterizations rely (Section \ref{results_decompose_improv}).
We also briefly discuss the results for the deep tropics (Section \ref{results_tropics})
Lastly,  in Section \ref{conclusion} we give a summary and the conclusions of the study.

%
%
 \section{Data and Methods}
 \label{data_methods}

 \subsection{Simulations}
 \label{data_methods_simulation}
 
The high resolution data was obtained from a quasi-global aquaplanet simulation (referred to as hi-res) on an equatorial beta plane using the System for Atmospheric Modeling (SAM) version 6.3 \cite{SAM_paper}. The domain of the simulation has a meridional extent of $17,280$ km and a zonal width of $6,912$ km, equivalent to a latitude range from $-78.5^\circ$ to $78.5^\circ$ and a longitudinal extent of 62.2$^\circ$ at the equator. To reduce the computational resources necessary to run a quasi-global simulation that resolves deep convection, we use a horizontal grid spacing of 12 km combined with a hypohydrostatic-rescaling factor of 4. Hypohydrostatic rescaling increases the horizontal length scale of convection without affecting the larger-scale flow \cite{kuang2005new,boos2016convective,fedorov2019tropical}. The sea surface temperature is specified to be the ``qobs'' distribution of \citeA{qobSST}, which is  zonally and hemispherically symmetric and peaks at the equator. The default time step is 24 seconds, but this time step is reduced if the  Courant-Friedrichs-Lewy (CFL) condition would otherwise be violated.  There are 48 vertical layers that extend up to 28.7 km.  The first $100$ days are considered as spinup, and we use three-dimensional instantaneous snapshots from a subsequent $337.5$ days that were saved every 3 model hours.  Detailed description of this simulation can be found in \citeA{Yani_natcommun}.

 To obtain training data, we follow the coarse-graining protocol described in \citeA{Yani_2021GRL}. In short, for each three-dimensional snapshot from hi-res, we coarse grain the prognostic variables as well as the temperature, vertical advective fluxes of energy, non-precipitating water and momentum, tendency of precipitating water due to cloud microphysics, turbulent diffusivity and radiative heating. Coarse-graining is performed by a spatial averaging to a horizontal grid spacing of 192 km (16$\times$16 grid boxes). Subgrid tendencies and fluxes are then calculated using the equations of the model. SAM uses an Arakawa C-grid \cite{arakawa1977computational} which introduces some challenges regarding how to coarse grain variables that are not found on the same horizontal grid (see discussion in the Supporting Information of \citeA{Yani_momentum_flux}). Here we choose to coarse grain the data such that the coarse-grained data is found on a collocated grid (see Figure S1 in \citeA{Yani_momentum_flux}). In other words, the coarse-grained quantities, including prognostic variables, tendencies, and fluxes, are all on the same horizontal grid.
%

 \subsection{Neural-Network Parameterizations}
 \label{data_methods_NN}
 
\begin{figure}
\centering
\includegraphics[width=1.0\linewidth]{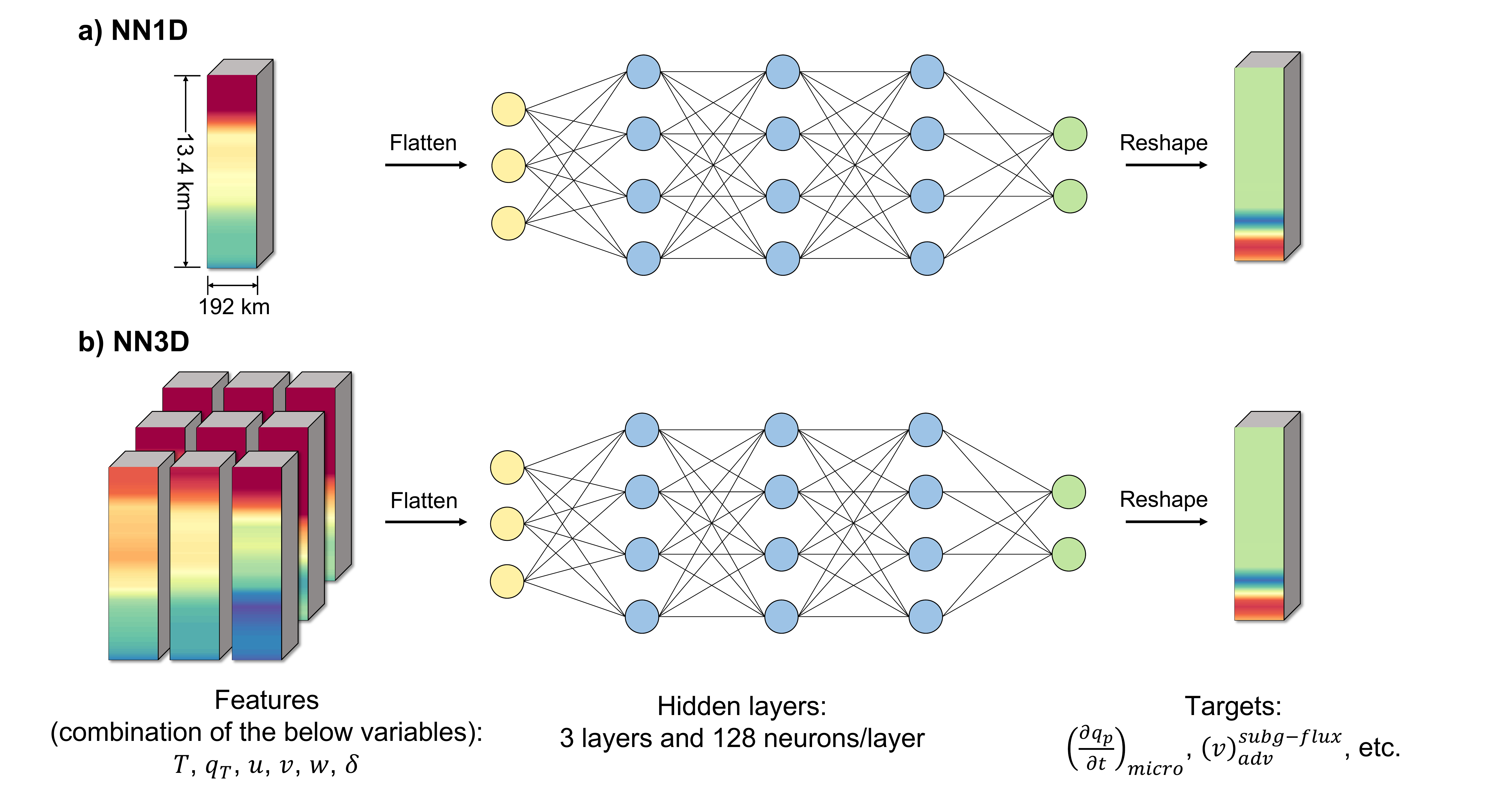}
\caption{Schematic of the neural-network (NN) architectures we use in this study. The single-column parameterization (NN1D) uses inputs taken from a single atmospheric column that is the same as the target column containing the outputs, whereas the non-local parameterization (NN3D) uses inputs taken from the 3$\times$3 columns of data centered on the target column. A combination of the following atmospheric variables are used as inputs: temperature ($T$), non-precipitating water mixing ratio ($q_T$), zonal, meridional and vertical velocities ($u, v, w$), and the horizontal wind divergence ($\delta$), but the schematic only shows one input variable for illustration purposes. The targets are single-columns of eight output variables, including the tendency of total precipitating water mixing ratio due to microphysics ($\left( {\partial q_p}/{\partial t} \right)_{\rm{micro}}$) and the subgrid meridional momentum flux due to vertical advection ($\left( v \right)^{\rm{subg-flux}}_{\rm{adv}}$).}
\label{NN_schematics}
\end{figure}

We train both a single-column parameterization (NN1D) and a non-local parameterization (NN3D), where the non-local parameterization uses inputs  from the 3$\times$3 atmospheric columns on the coarse grid (Figure~\ref{NN_schematics}).
 The atmospheric variables that are used as the default inputs for NN1D and NN3D are the coarse-grained absolute temperature ($T$), non-precipitating water mixing ratio (sum of water vapor, cloud liquid water and cloud ice mixing ratios; $q_{\rm{T}}$), and zonal ($u$) and meridional ($v$) velocities.
 As discussed in the introduction, we also train single-column NN parameterizations that use the above four local input features as well as an additional input of the local vertical wind (this NN is referred to as NN1D+$w$) or the horizontal wind divergence (this NN is referred to as NN1D+$\delta$ where $\delta = \frac{\partial u}{\partial x}+\frac{\partial v}{\partial y}$).
 From symmetry considerations, in the Southern Hemisphere the meridional coordinate is flipped and the meridional velocity is changed in sign before training so that the NNs we train can generalize across hemispheres. 
 The number of inputs is 120 for NN1D (4 feature variables, each variable with 30 vertical levels), 1,080 for NN3D (4 feature variables, each variable with 30 vertical levels and 3$\times$3 columns) and 150 for NN1D+$w$ and NN1D+$\delta$ (5 feature variables, each variable with 30 vertical levels).

We train NNs to predict the following eight target variables simultaneously: the tendency of total precipitating water mixing ratio due to microphysics ($\left( {\partial q_p}/{\partial t} \right)_{\rm{micro}}$); the subgrid fluxes of total non-precipitating mixing ratio due to vertical advection ($\left( q_T \right)^{\rm{subg-flux}}_{\rm{adv}}$) and sedimentation ($\left( q_T \right)^{\rm{subg-flux}}_{\rm{sed}}$); the tendency due to radiation ($\left( {\partial H_L}/{\partial t} \right)_{\rm{rad}}$); the subgrid energy flux due to vertical advection ($\left( H_L \right)^{\rm{subg-flux}}_{\rm{adv}}$); the coarse-grained turbulent diffusivity ($\overline{D}$); and the subgrid zonal and meridional momentum flux due to vertical advection ($\left( u \right)^{\rm{subg-flux}}_{\rm{adv}}$ and $\left( v \right)^{\rm{subg-flux}}_{\rm{adv}}$). A detailed description of how the outputs are calculated is found in \citeA{Yani_2021GRL}, except for the subgrid momentum fluxes which are described in \citeA{Yani_momentum_flux}.

 For NN3D the outputs are only predicted at the center column, such that both local and non-local NNs have the exact same outputs. We only use inputs and outputs from the lowest 30 vertical levels of the model (extending to 13.4 km) to prevent the NNs from predicting near the sponge layer which is active above $20$ km, and also because previous studies suggest that using stratospheric information  can lead to numerical instabilities when parameterizations are implemented online \cite{Brenowitz_2019,Yani_2021GRL}.

 We use 3-hourly snapshots taken from 337.5 days, resulting in 2,700 time snapshots. The coarse-grained data contains 90$\times$36 atmospheric columns (samples) for each snapshot, but the two outermost columns of data in each snapshot are not used to simplify both the training of the non-local parameterization and the calculation of the horizontal divergence which is used later. Therefore, each time snapshot contains (90-4)$\times$(36-4)=2,752 samples, resulting in total of 7,430,400 samples.
 We use the first 50\% of the simulated data for training (3,715,200 samples), the middle 10\% of the simulated data for validation (743,040 samples) and the remaining 40\% of the simulated data for testing (2,972,160 samples). All the results shown in this paper are based only on the testing set. 
The reason for using only 50\% of data for training is because we want to have a large testing dataset to ensure robust results for case composites. We verified that training on 80\% of the data leads to similar coefficient of determination (R$^2$) values compared to when training on 50\% of the data (Figure S1).

Before training, the input variables are standardized such that each input at each vertical level has a mean of zero and a standard deviation of one.
 The outputs are also standardized by removing the mean and rescaling by the standard deviation, but the mean and standard deviation are calculated over all vertical levels, such that the output standardization consists of a single mean value and a single standard deviation value for all levels of each output variable. 
 We use the mean squared error as a loss function, and we use the Adam optimizer \cite{Adam} to update the weights and biases.
 The training process is the same for all networks we train. The NNs are first trained for 7 epochs, with a cyclic learning rate \cite{smith2017cyclical} bounded by $2\times10^{-4}$ to $2\times10^{-3}$. They are then trained for another 5 epochs with reduced cyclic learning rate bounded by $2\times10^{-5}$ to $2\times10^{-4}$.
 We apply ``early stopping'' by using the validation data to evaluate the network after each epoch, and we choose the NN weights and biases that performed best on the validation data. The default NN architecture we use in this study has 3 hidden layers where each layer has 128 neurons, and we use rectified linear unit activations (ReLu) except in the output layer. Figure S2 shows the training and validation loss versus epoch for NN1D, NN3D, NN3D+$\delta$ and NN1D+$w$, as well as the learning rate during each epoch.

 \subsection{Layer-wise Relevance Propagation}
 \label{data_methods_LRP}

 We use layer-wise relevance propagation (LRP; \citeA{Batch_LRP}) to better understand the inputs that are most important for NN3D. LRP propagates the relevance score from the output layer back to the input layer. Therefore, each input has an associated relevance score, where a higher relevance score indicates that the NN relies more on this input for the specific sample that is tested. The propagation rules used are given in Text S1.

 Different output variables with distinct physical processes can rely on different inputs. The relevance score from the multi-target NN might be hard to interpret because it entangles the physical processes from all the eight target variables. Therefore, when appyling LRP for a given output variable, we retrain the NN parameterizations to have only the single output variable we are interested in.
 LRP provides the relevance score for each input variable and grid box at 30 vertical levels. Because we want to focus on the reliance of NNs on horizontally non-local inputs which is the novelty of this paper,  we sum the relevance across heights, treating height levels as color channels in image recognition problems. However, the relevance score can be positive or negative, and both signs in relevance are physically meaningful. Therefore, we sum over the absolute value of the relevance score across height levels. We then normalize the vertically summed relevance sample by sample, such that the relevance values for all variables and columns add up to one in each sample. This better illustrates the relative importance of each variable for each column, and the magnitude of vertically summed relevance can be easily interpreted. 
 
%
%
 \section{Results}
 \label{results}

 \subsection{Performance of parameterizations}
 \label{results_identify_improv}
 
\begin{figure}
 \centering
 \includegraphics[width=1.0\linewidth]{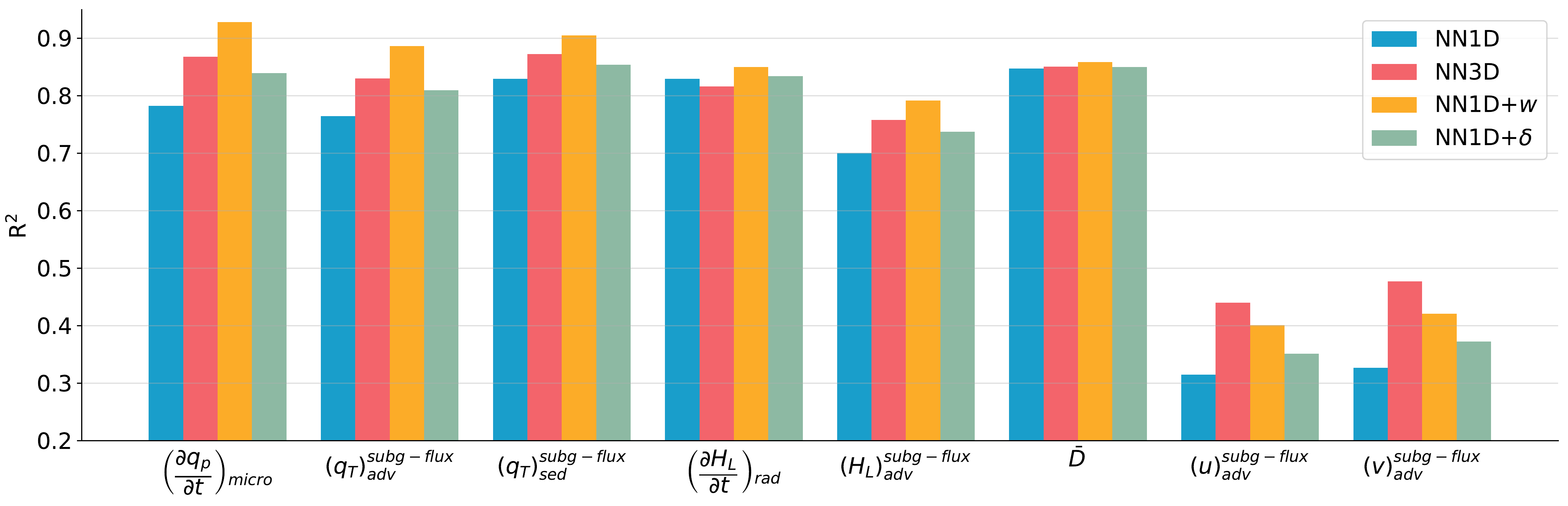}
 \caption{Global coefficient of determination (R$^2$) for a variety of output variables using a single-column parameterization (NN1D; blue), a non-local parameterization (NN3D; red), a single-column parameterization with the vertical wind ($w$) being an additional input feature (NN1D+$w$; yellow), and a single-column parameterization with the horizontal wind divergence ($\delta$) being an additional input feature (NN1D+$\delta$; green).
 All four NNs predict all the presented variables simultaneously (multi-output prediction). The predicted outputs are the tendency of total precipitating water mixing ratio due to microphysics ($\left( {\partial q_p}/{\partial t} \right)_{\rm{micro}}$); the subgrid flux of total non-precipitating mixing ratio due to vertical advection ($\left( q_T \right)^{\rm{subg-flux}}_{\rm{adv}}$) and sedimentation ($\left( q_T \right)^{\rm{subg-flux}}_{\rm{sed}}$); the subgrid tendency due to radiation ($\left( {\partial H_L}/{\partial t} \right)_{\rm{rad}}$); the subgrid energy flux due to vertical advection ($\left( H_L \right)^{\rm{subg-flux}}_{\rm{adv}}$); the coarse-grained diffusivity ($\overline{D}$); and the subgrid zonal and meridional momentum flux due to vertical advection ($\left( u \right)^{\rm{subg-flux}}_{\rm{adv}}$ and $\left( v \right)^{\rm{subg-flux}}_{\rm{adv}}$).}
 \label{R2_more_targets}
 \end{figure}

To measure the performance of NN1D, NN3D,  NN1D+$w$ and NN1D+$\delta$, we calculate the global R$^2$ values by concatenating vertical columns for each target variable (Figure~\ref{R2_more_targets}). The non-local parameterization NN3D improves on NN1D for almost all variables. The only exceptions are the tendency due to radiation and the coarse grained diffusivity for which the non-local parameterization is either a disimprovement or unaltered, respectively. For most output variables, NN1D+$w$ has better performance than NN3D, with the important exception that NN3D outperforms NN1D+$w$ for the zonal and meridional subgrid momentum fluxes. As discussed in the introduction, it is not clear whether the vertical velocity should be included as an input from the point of view of causality and robustness in online simulations, and thus we present results with and without the vertical velocity as an input.
 
 The performance of NN1D+$\delta$ is better than NN1D but worse than the performance of NN1D+$w$ and NN3D, and the relation of  the horizontal wind divergence ($\delta$) to the non-local wind inputs in NN3D will be discussed further in Section \ref{results_decompose_improv}. It may seem surprising that using the horizontal wind divergence as an input is not equivalent to using the coarse-grained vertical velocity as an input given that these are related by the anelastic mass continuity equation in SAM. However, the coarse-grained (i.e., horizontal averaged) vertical velocity is directly related by mass continuity to line averages of the horizontal winds on the boundaries of the grid cells. These line averages are effectively subgrid compared to the coarse-grained horizontal winds. Thus, including the vertical velocity as an input is not equivalent to including the horizontal divergence as an input, and we present separate results for parameterizations using these inputs.

 One caveat of the global R$^2$ values is that they could be overstated because they partly reflect the NN parameterization correctly predicting the mean at each vertical level.
To verify that the NN parameterizations predict accurately beyond the means at each vertical level, we also tested the performance  after the means at each vertical level are first removed (Figure~S3). 
When the means at each vertical level are first removed, R$^2$ values for all the targets are slightly lower (and substantially so for the turbulent diffusivity), but the differences between different NNs remain similar.
 
 From now on, we primarily focus on two output variables: the tendency of total precipitating water mixing ratio due to microphysics ($\left( {\partial q_p}/{\partial t} \right)_{\rm{micro}}$) and the subgrid meridional momentum flux due to vertical advection ($\left( v \right)^{\rm{subg-flux}}_{\rm{adv}}$).
  We choose to focus on these two output variables because (a) their prediction is most accurate for NNs that use different inputs, (b) $\left( {\partial q_p}/{\partial t} \right)_{\rm{micro}}$ is directly related to surface precipitation (Text~S2) which is a variable of great interest, and (c) subgrid momentum transport has previously been found to be challenging to predict \cite{Yani_momentum_flux}, and therefore it is especially interesting to investigate how its prediction can be improved using non-local inputs. The results for $\left( u \right)^{\rm{subg-flux}}_{\rm{adv}}$ are similar in most regards to the results for $\left( v \right)^{\rm{subg-flux}}_{\rm{adv}}$, but the improvement from using non-local inputs is slightly greater for $\left( v \right)^{\rm{subg-flux}}_{\rm{adv}}$, and thus we focus on it throughout the paper.

 \begin{figure}
 \centering
 \includegraphics[width=1.0\linewidth]{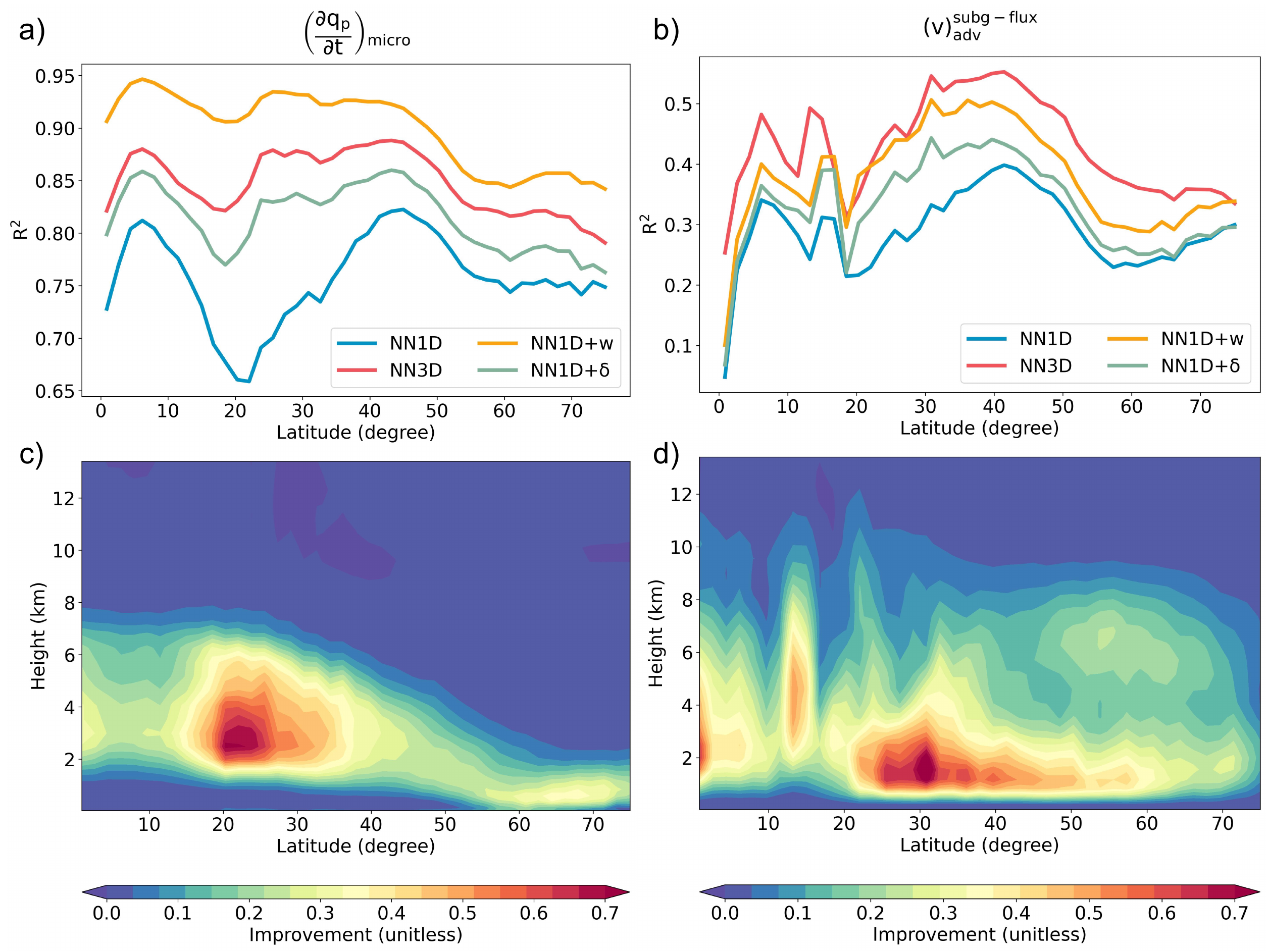}
 \caption{\textbf{(a-b)} Coefficient of determination (R$^2$) as a function of latitude for (a) the precipitating water tendency due to microphysical processes ($\left( {\partial q_p}/{\partial t} \right)_{\rm{micro}}$) and (b) the subgrid meridional momentum flux due to vertical advection ($\left( v \right)^{\rm{subg-flux}}_{\rm{adv}}$)  for different neural networks (NNs): a single-column parameterization (NN1D; blue), a single-column parameterization with an additional 1D vertical velocity input (NN1D+$w$; yellow), a single-column parameterization with an additional 1D wind divergence input (NN1D+$\delta$; green), and a non-local parameterization (NN3D; red).
 \textbf{(c-d)} The zonal-mean improvement index (equation~\ref{eqn_improv}) for NN3D compared to NN1D averaged over all test samples as a function of latitude and height for (c) $\left( {\partial q_p}/{\partial t} \right)_{\rm{micro}}$  and (d) $\left( v \right)^{\rm{subg-flux}}_{\rm{adv}}$.}
 \label{R2_and_improv}
 \end{figure}

 We next consider $R^2$ values calculated at each latitude for the two output variables of interest (Figure~\ref{R2_and_improv}a,b). 
 Both NN3D and NN1D+$w$ perform substantially than NN1D at all latitudes, where NN1D+$w$ predicts more accurately  $\left( {\partial q_{\rm{p}}}/{\partial t} \right)_{\rm{micro}}$, and NN3D predicts more accurately  $\left( v \right)^{\rm{subg-flux}}_{\rm{adv}}$.
Unsurprisingly given the close relation of $\left( {\partial q_{\rm{p}}}/{\partial t} \right)_{\rm{micro}}$ to precipitation rates, both NN3D and NN1D+$w$ outperform NN1D in predicting extreme precipitation (Figure S4).

 To verify that the improvement is robust to changes in the network architecture, and is not due to the larger number of tunable parameters used in NN3D and NN1D+$w$ compared to NN1D, we also train networks with different complexities by adjusting the number of hidden layers and neurons. Specifically, we compare the default architecture of 3 hidden layers and 128 neurons per layer to a shallow network with a single hidden layer and 64 neurons, and a more complex architecture with 8 hidden layers and 256 neurons per layer.  In all cases, performance improves when changing from a single hidden layer to 3 hidden layers, but performance stays similar or disimproves when changing from 3 hidden layers to a deeper network with 8 hidden layers (Figure S5). More importantly we find that NN3D or NN1D+$w$ even with a single hidden layer outperforms NN1D for all architectures. This implies that the additional information provided as inputs to NN3D and NN1D+$w$ improves their performance. For the rest of the paper, we will only focus on the results using 3 hidden layers and 128 neurons per layer.

\subsection{In which atmospheric states does non-locality help the parameterization in midlatitudes?}
 \label{results_explain_improv}

 Next, we want to understand for which atmospheric states NN3D predicts better than NN1D. We first define an improvement index for each individual test sample for each of the output variables and each atmospheric level as:
 \begin{linenomath*}
 \begin{equation}
   \textrm{Improvement} = \frac{ \left(\textrm{NN1D} - \textrm{true}\right)^2 - \left(\textrm{NN3D} - \textrm{true}\right)^2}{\sigma_{true}^2},
 \label{eqn_improv}
 \end{equation}
 \end{linenomath*}
 where $\left(\textrm{NN1D} - \textrm{true}\right)^2$ is the squared error of NN1D output, $\left(\textrm{NN3D} - \textrm{true}\right)^2$ is the squared error of NN3D output, and $\sigma_{true}^2$ is the variance of the ground truth over the column and over all testing samples which is a latitude-dependent variable. 
 A large improvement index indicates that  the difference between the squared errors of NN1D and NN3D is large compared to the climatological variance at a given latitude.
 By taking the zonal mean of the improvement indices at every altitude and latitude, we find that most of the improvements occur in subtropical and mid-latitude tropospheric regions for both $\left( {\partial q_{\rm{p}}}/{\partial t} \right)_{\rm{micro}}$ and $\left( v \right)^{\rm{subg-flux}}_{\rm{adv}}$, although the improvements in $\left( v \right)^{\rm{subg-flux}}_{\rm{adv}}$ are more spread in the vertical and maximize at lower altitudes (Figure \ref{R2_and_improv}c,d).
 We note that for $\left( {\partial q_{\rm{p}}}/{\partial t} \right)_{\rm{micro}}$, the absolute improvement in the tropics is even larger than the absolute improvement in the mid-latitudes, but the variance in the tropics is very large, and therefore the relative improvement is smaller in that region. For $\left( v \right)^{\rm{subg-flux}}_{\rm{adv}}$, the biggest improvement occurs in the mid-latitudes in the both absolute and relative sense.

We will focus on the midlatitude band 20$^{\circ}$ -- 40$^{\circ}$ in both hemispheres which shows large improvements for NN3D. The region of greatest improvement actually extends somewhat further equatorward than $20^\circ$ latitude, but we focus on 20$^{\circ}$ -- 40$^{\circ}$ to avoid mixing the tropical and mid-latitude dynamical regimes.  
We next compare groups of cases
in this latitude band with large and small improvement indices. Each case is a testing sample with 3$\times$3 horizontal grid boxes and 30 vertical levels, and thus has a horizontal length scale of 574 km and a vertical height scale of 13.4 km. We find that cases with large improvement tend to have heavier precipitation on average than cases with no improvement. 
Therefore, in order to make a fair comparison between cases with and without improvement, we only select cases that have true instantaneous precipitation between 50 -- 70 mm day$^{-1}$ (corresponding to 99.5\textsuperscript{th} -- 99.9\textsuperscript{th} percentile of precipitation over the mid-latitude band). 
From cases with these precipitation rates, we use the improvement criterion (equation~\ref{eqn_improv}) to select the 300 cases for each output variable that have the largest column-mean improvement index, which we refer to as the ``cases with largest improvement''. We also select the 300 cases with the smallest column-mean improvement index that is still positive, which we refer to as the ``cases with little improvement''. 
Among ``cases with largest improvement'', the improvement indices range between 11.5 -- 92.7 for $\left( {\partial q_{\rm{p}}}/{\partial t} \right)_{\rm{micro}}$, and 16.6 -- 511.9 for $\left( v \right)^{\rm{subg-flux}}_{\rm{adv}}$. And among ``cases with little improvement'', the improvement indices range between $2.2\times10^{-3}$ -- 0.6 for $\left( {\partial q_{\rm{p}}}/{\partial t} \right)_{\rm{micro}}$, and $1.5\times10^{-3}$ -- 0.6 for $\left( v \right)^{\rm{subg-flux}}_{\rm{adv}}$.
41\% of the testing samples have negative improvement indices for $\left( {\partial q_{\rm{p}}}/{\partial t} \right)_{\rm{micro}}$ and 46\% for $\left( v \right)^{\rm{subg-flux}}_{\rm{adv}}$, but most of the negative improvements are close to zero.


 \begin{figure}
 \centering
 \includegraphics[width=1.0\linewidth]{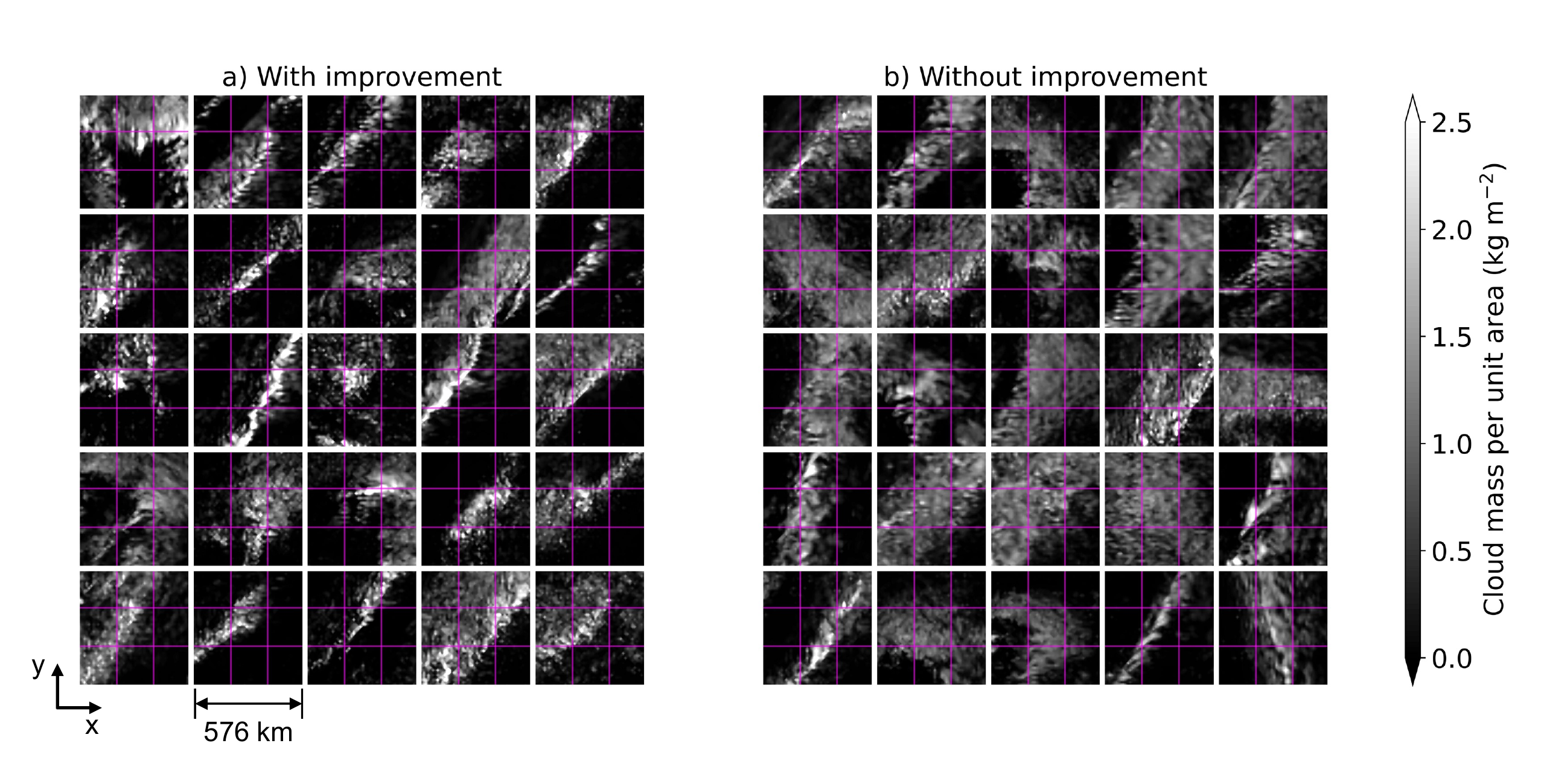}
 \caption{Snapshots of the column-integrated cloud mass per unit area plotted from the hi-res output for mid-latitude cases with and without improvement when using the non-local parameterization NN3D to predict $\left( {\partial q_{\rm{p}}}/{\partial t} \right)_{\rm{micro}}$ (similar results are found for $\left( v \right)^{\rm{subg-flux}}_{\rm{adv}}$ as shown in Figure S6).  
 Shown are  25 randomly-chosen cases from \textbf{(a)} the 300-member group with large-improvement when the non-local parameterization is used, and \textbf{(b)} the 300-member group with little improvement when the non-local parameterization is used.
 See Section \ref{results_explain_improv} for details on how these groups were chosen which involves only selecting cases with heavy precipitation in the latitude band 20$^{\circ}$ -- 40$^{\circ}$. The domains are equivalent to 3$\times$3 coarse-grained grid boxes (576 km in each horizontal direction) with the edges of the coarse-grained grid boxes plotted in magenta.}
 \label{QN_postage}
 \end{figure}

 We find that cases where NN3D improves the prediction tend to have different cloud shapes compared to the cases where NN3D does not improve the prediction (Figure~\ref{QN_postage} for $\left( {\partial q_{\rm{p}}}/{\partial t} \right)_{\rm{micro}}$ and Figure S6 for $\left( v \right)^{\rm{subg-flux}}_{\rm{adv}}$). Specifically, for cases with improvement the clouds tend to be organized in coherent linear and narrow features which are squall lines associated with atmospheric fronts.
 Coarse-graining smears out sharp boundaries such as fronts and narrow convective features such as squall lines making the prediction of subgrid tendencies and fluxes more difficult. Therefore single-column coarse-grid inputs might not be informative enough to make an accurate prediction when such strong subgrid variability is present, but a non-local parameterization may be able to use non-local information in order to better understand the atmospheric conditions and improve the prediction of the effect of subgrid processes on the resolved scales. For cases without improvement, clouds are more uniformly distributed, and the inputs from non-local columns does not convey information which assists in the prediction of the effect of subgrid processes. 
 We find that doing a similar analysis for NN1D+$w$ gives similar results (e.g., 260 cases out of the 300 improved cases are shared between NN3D and  NN1D+$w$ for $\left( {\partial q_{\rm{p}}}/{\partial t} \right)_{\rm{micro}}$).

 \begin{figure}
 \centering
 \includegraphics[width=0.7\linewidth]{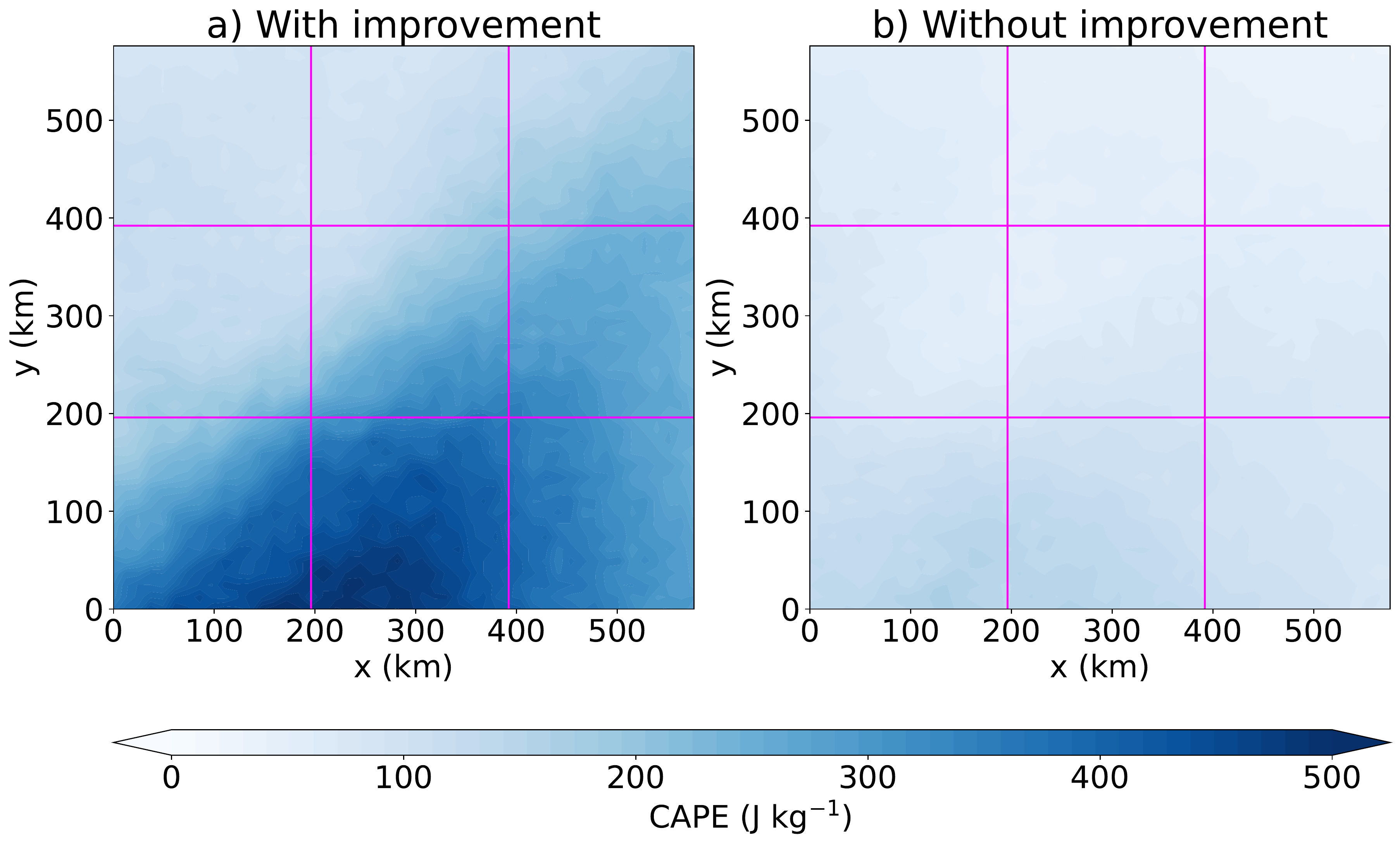}
 \caption{Composite of convective available potential energy (CAPE) for cases in the mid-latitudes (20$^{\circ}$ -- 40$^{\circ}$) 
 for \textbf{(a)} the group of 300 cases with largest improvement when the non-local parameterization NN3D is used to predict
 $\left( {\partial q_{\rm{p}}}/{\partial t} \right)_{\rm{micro}}$,
and \textbf{(b)} the group of 300 cases with little improvement in predicting  $\left( {\partial q_{\rm{p}}}/{\partial t} \right)_{\rm{micro}}$.
 Each case is equivalent to a single testing sample, and see Section \ref{results_explain_improv} for details of how the groups of cases are chosen. CAPE is calculated from the hi-res output. The domain shown is equivalent to 3$\times$3 coarse-grained grid boxes (576 km in each horizontal direction) with the edges of the coarse-grained grid boxes plotted in magenta. Similar results for $\left( v \right)^{\rm{subg-flux}}_{\rm{adv}}$ are shown in Figure S7.}
 \label{Midlat_CAPE}
 \end{figure}

 To better understand in which atmospheric conditions a non-local parameterization is superior to  a single-column parameterization, we calculate the convective available potential energy (CAPE) from the high resolution data for cases with largest improvement and cases with little improvement. CAPE is calculated assuming reversible ascent with convective inhibition (CIN) being removed as described in Text S3 following \citeA{Muller_CAPE}.  We find that  the CAPE  is larger  in cases with the largest improvement compared to cases with little improvement (Figure~\ref{Midlat_CAPE} for $\left( {\partial q_{\rm{p}}}/{\partial t} \right)_{\rm{micro}}$, and similar results in Figure S7 for $\left( v \right)^{\rm{subg-flux}}_{\rm{adv}}$), suggesting the atmosphere is more unstable to convection in the improved cases. The CAPE composites show higher CAPE at the equatorward and eastward ends of the 3$\times$3 subdomain, and this is because of the general increase in CAPE equatorward and the structure of the flow for these cases which are all precipitating. 
 We also find that the improved cases are more unstable for two other measures of instability as shown in Figure S8-S9: the saturation potential vorticity (negative values are indicative of conditional symmetric instability which could give rise to slantwise convection) and the vertical derivative of the saturation moist static energy (negative values are indicative of upright conditional instability). 
 Calculation of these instability metrics is described in Text S4. One possibility for the improvement in non-local parameterization is that it could be especially relevant for correctly predicting slantwise convection since slantwise convection is not a purely vertical process and is sensitive to horizontal gradients that can be estimated by the non-local parameterization, and since slantwise convection may also involves more than one coarse atmospheric column. Interestingly, the latitude band we find that has the greatest relative improvement (roughly $10^\circ -40^\circ$) is similar to the latitude band in which conditional symmetric instability is most favored over upright convective instability in reanalysis data (see Figure 2b of \citeA{Chen_MPV}). However, we note that NN3D being better at recognizing conditional symmetric instability is only one possible reason for the improvement, and further work is needed to support this hypothesis.

 \subsection{Which non-local inputs are useful in midlatitudes and how do they relate to horizontal wind divergence?}
 \label{results_decompose_improv}
 
 We first test whether extending the non-locality beyond using information from the closest neighbours (3$\times$3 atmospheric columns) gives further improvements. To do this, we trained NNs with non-local information from $5\times5$ and $7\times7$ atmospheric columns (Figure~S10).
For almost all output variables, the best performing networks rely on $3\times3$ atmospheric columns. We conclude that extending the non-locality beyond the closest neighbouring atmospheric columns is not helpful to further improve the parameterization.

 \begin{figure}
 \centering
 \includegraphics[width=\linewidth]{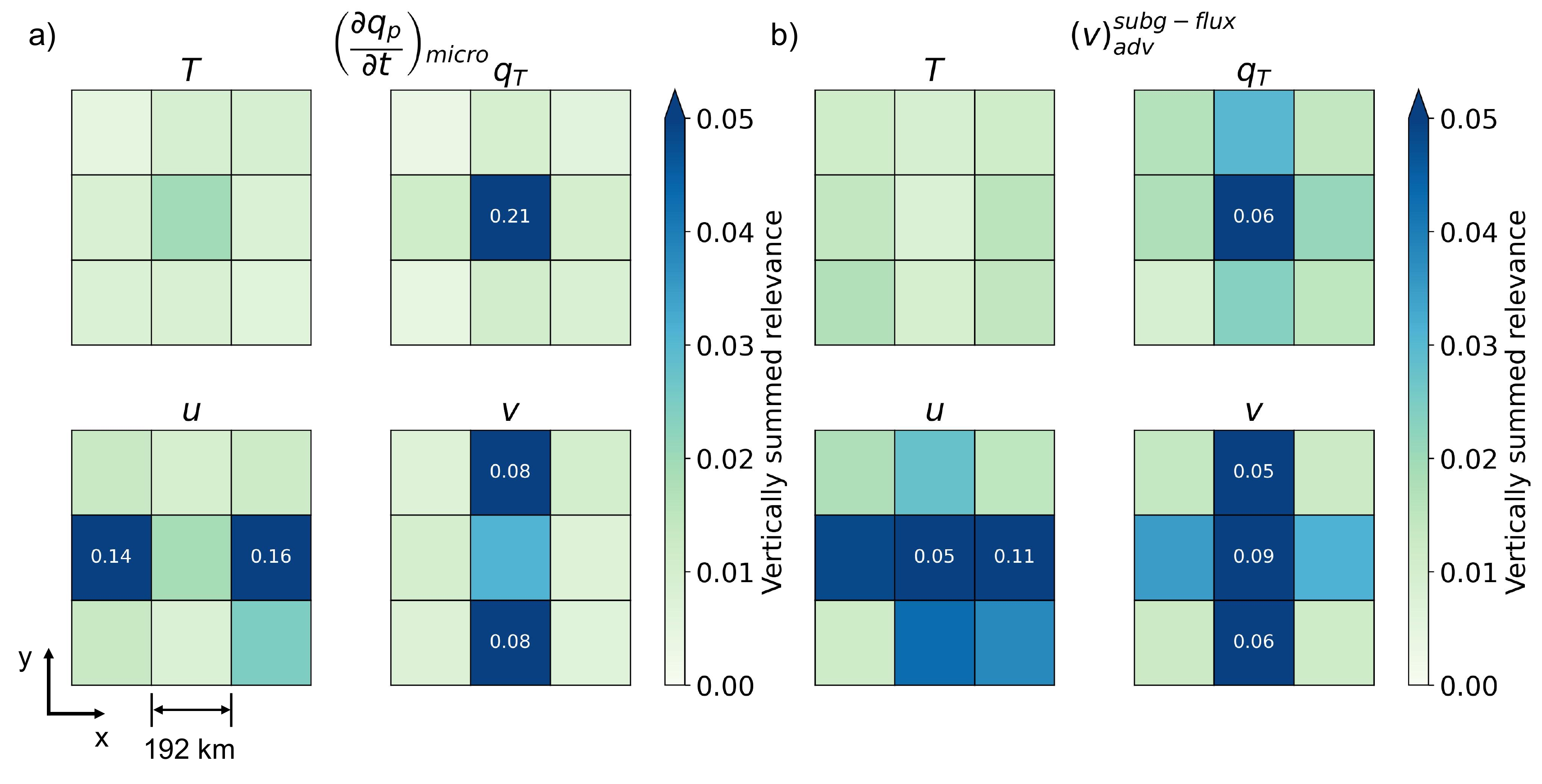}
 \caption{Vertically summed relevance (as calculated from layer-wise relevance propagation) averaged over the group of 300 mid-latitude (20$^{\circ}$ -- 40$^{\circ}$) cases with largest improvement when the non-local parameterization NN3D is used to predict \textbf{(a)} $\left( {\partial q_{\rm{p}}}/{\partial t} \right)_{\rm{micro}}$ and \textbf{(b)} $\left( v \right)^{\rm{subg-flux}}_{\rm{adv}}$. Each sub-panel shows the vertically summed relevance for the 3$\times$3 atmospheric input columns of a different input variable: temperature ($T$), non-precipitating water mixing ratio ($q_T$), zonal wind ($u$), and meridional wind ($v$).  To better illustrate the variations in relevance, we set the upper limit of the colorbar to 0.05 and explicitly write the relevance for the inputs that saturate the color bar. See Section \ref{results_explain_improv} for details of how the cases were chosen.}
 \label{Midlat_LRP_NN3D}
 \end{figure}

 We next use LRP to calculate the relevance score for the cases showing the largest improvements from NN3D (Figure~\ref{Midlat_LRP_NN3D}). A separate NN3D is trained to predict each of the two output variables on their own in order to perform the LRP analysis with only one output variable. This is to prevent mixing the relevance for different physical processes that matter for different target variables. The groups of cases with largest improvement are the same groups of cases selected in previous sections for consistency. Each grid box in Figure \ref{Midlat_LRP_NN3D} indicates a single column of data for a different input variable, with the absolute value of the relevance score being vertically summed and then averaged across cases. 
 Figure~S11 shows that the LRP results are robust to changes in the LRP parameters (described in Text S1).
 
 Starting with the NN3D predicting $\left( {\partial q_{\rm{p}}}/{\partial t} \right)_{\rm{micro}}$, we find that for the temperature and non-precipitating water input variables, the most relevant grid box is the center column, which means that thermodynamic and moisture variables follow the reasoning of using a single column for parameterizations (Figure~\ref{Midlat_LRP_NN3D}a). However, the most relevant grid boxes for the wind input variables are found at the non-local atmospheric columns. 
 Interestingly, for the zonal wind the most relevant columns are east and west of the center column, whereas for the meridional wind the most relevant columns are north and south of the center column, and the relevance is symmetric about the center column, which suggests that the NN is using the non-local winds to reconstruct the horizontal wind divergence $\delta = \frac{\partial u}{\partial x}+\frac{\partial v}{\partial y}$.  For the NN3D predicting $\left( v \right)^{\rm{subg-flux}}_{\rm{adv}}$,  the local moisture input is still important, but non-local moisture inputs also contribute (Figure \ref{Midlat_LRP_NN3D}b). Furthermore, NN3D relies heavily on both the local and non-local columns of horizontal wind fields. Unlike for the NN3D that predicts $\left( {\partial q_{\rm{p}}}/{\partial t} \right)_{\rm{micro}}$, the relevance for NN3D that predicts $\left( v \right)^{\rm{subg-flux}}_{\rm{adv}}$  is not symmetric around the center column, and there are high-relevance regions that are not necessary to calculate the horizontal wind divergence, which implies that this NN uses features beyond the horizontal wind divergence.
  Overall, these patterns imply that NN3D is likely reconstructing the horizontal wind divergence from non-local wind fields to help predict $\left( {\partial q_{\rm{p}}}/{\partial t} \right)_{\rm{micro}}$. This conclusion from LRP is consistent with the increase in performance when the divergence is added as an input to NN1D (Figure \ref{R2_and_improv}a). The horizontal wind divergence may also be reconstructed to help predict $\left( v \right)^{\rm{subg-flux}}_{\rm{adv}}$ but other aspects of the non-local winds are also being used for that output.

\begin{figure}
 \centering
 \includegraphics[width=\linewidth]{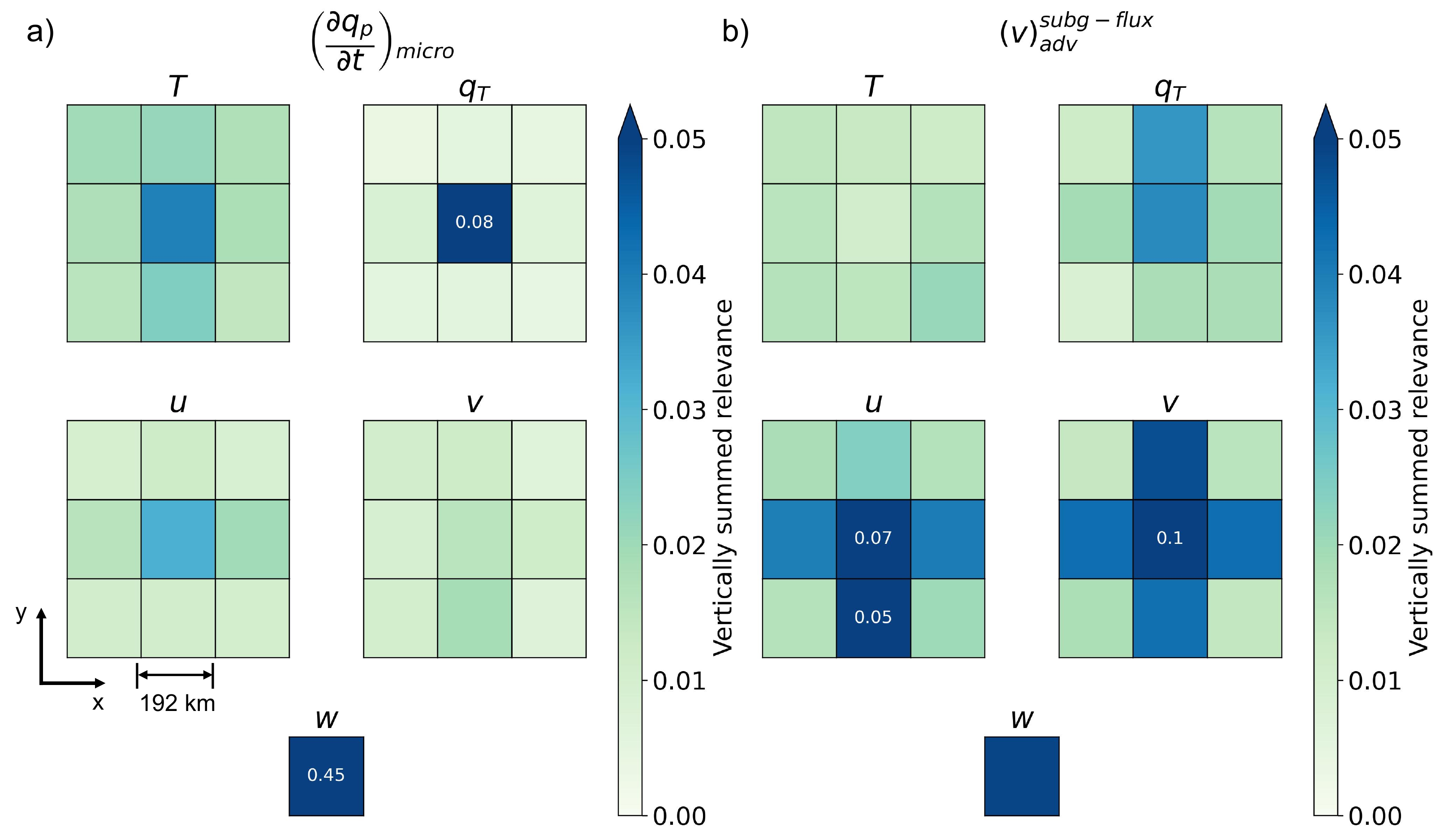}
 \caption{As in Figure \ref{Midlat_LRP_NN3D}, but the relevance is plotted for non-local NN parameterizations that include the vertical velocity at the center column as an additional input (NN3D+$w$) for cases where there are large improvements for NN3D+$w$ compared to NN1D.} 
 \label{Midlat_LRP_NN3D_w}
 \end{figure}

We next investigate the relevance scores for cases with largest improvement for 
 a non-local NN that additionally gets as an input the vertical wind at the center column (this network is referred to as NN3D+$w$).
 LRP results for NN3D+$w$ could potentially show the competing effects between local vertical velocity and non-local horizontal wind fields if the non-local winds are used to construct the horizontal divergence and thus approximate the vertical wind.
 For NN3D+$w$ that predicts $\left( {\partial q_{\rm{p}}}/{\partial t} \right)_{\rm{micro}}$, we find that the relevance for non-local wind inputs reduces dramatically compared to NN3D, and 45 $\%$ of the total relevance score comes from the vertical wind at the center column (Figure~\ref{Midlat_LRP_NN3D_w}a). This result provides evidence that NN3D tries to estimate the vertical wind through the horizontal wind divergence in order to better
 predict $\left( {\partial q_{\rm{p}}}/{\partial t} \right)_{\rm{micro}}$,
 and hardly relies on non-local information beyond the horizontal wind divergence.
 However, for NN3D+$w$ that predicts $\left( v \right)^{\rm{subg-flux}}_{\rm{adv}}$, we find that the relevance for the non-local wind inputs reduces only by a moderate amount
(Figure~\ref{Midlat_LRP_NN3D_w}b). 
Furthermore, only $4.7\%$ of the total relevance score comes from the vertical wind at the center column. These LRP results provide evidence that when NN3D is predicting
$\left( v \right)^{\rm{subg-flux}}_{\rm{adv}}$, it indeed relies on non-local information beyond that necessary to construct the horizontal wind divergence, and this is also consistent with the ability of NN3D to outperform NN1D+$w$ when predicting subgrid momentum transport (Figure~\ref{R2_and_improv}b).

 \subsection{Cases with improvement and role of non-local inputs in the tropics}
 \label{results_tropics}
 
 Following the same general approach as for the mid-latitudes, we select cases with largest improvement and cases with little improvement when using the non-local parameterization but for the deep tropics defined as latitudes between $10^\circ$S and $10^\circ$N. Cases with improvement are selected as the 300 cases with the largest column-mean improvement indices and with instantaneous precipitation between 100 -- 150 mm day$^{-1}$ (corresponding to the 99.5\textsuperscript{th} -- 99.9\textsuperscript{th} percentile in the deep tropics), resulting in column-mean improvement indices between 3.9 -- 35.6 for $\left( {\partial q_{\rm{p}}}/{\partial t} \right)_{\rm{micro}}$, and 14.3 -- 249.5 for $\left( v \right)^{\rm{subg-flux}}_{\rm{adv}}$. 
 Cases with little improvement are selected as the 300 cases with the smallest column-mean improvement indices that are still positive and with instantaneous precipitation between 100 -- 150 mm day$^{-1}$, resulting in column-mean improvement indices between 1.5$\times10^{-3}$ -- 0.3, and 1.3$\times10^{-3}$ -- 1.1 for $\left( {\partial q_{\rm{p}}}/{\partial t} \right)_{\rm{micro}}$ and $\left( v \right)^{\rm{subg-flux}}_{\rm{adv}}$ respectively.

 Although NN3D has better performance compared to NN1D at all latitudes, the cases showing improvement appear to be different in the deep tropics and mid-latitudes. 
 In the deep tropics, the cloud shapes for cases that have large improvements when using NN3D are not clearly distinguishable from cases that do not have an improvement as shown for example for the $\left( {\partial q_{\rm{p}}}/{\partial t} \right)_{\rm{micro}}$ output in Figure S12. Furthermore, 
 NN3D improves the prediction for more unstable conditions in the mid-latitudes (Figure~\ref{Midlat_CAPE}) but for indistinguishable or slightly more stable conditions in the tropics, as shown for example for cases with largest improvements for the $\left( {\partial q_{\rm{p}}}/{\partial t} \right)_{\rm{micro}}$ output in 
 Figure S13. However, LRP results suggest similar columns of non-local data are most relevant for both the tropics and mid-latitudes (compare Figures \ref{Midlat_LRP_NN3D} and S14). Thus, the atmospheric conditions under which improvement is found differs between the tropics and mid-latitudes, but similar non-local information is used in both regions.

%
%
 \section{Conclusions}
 \label{conclusion}

 In this study we show that a neural-net (NN) parameterization using inputs of temperatures, moisture and horizontal winds from  3$\times$3 atmospheric columns (NN3D; non-local in the horizontal) is generally more accurate compared to a single column NN parameterization (NN1D) in predicting the tendencies and fluxes due to different subgrid processes. 
 We find that the relative improvement is especially large for the subtropics and mid-latitudes.
 Cases in mid-latitudes with the largest improvements have heavy precipitation and squall-line feature associated with fronts, and they tend to be convectively and symmetrically more unstable than cases with little improvement.  
  We hypothesize that such unstable cases with large subgrid variations are more amenable to improvement through non-local information. 
 
 We use layer-wise relevance propagation, which is an interpretable machine-learning (ML) technique, to determine which non-local features the parameterizations relies on for its predictions.
 We focus on the prediction of two different target variables: the subgrid tendency of total precipitating water mixing ratio due to microphysics ($\left( {\partial q_p}/{\partial t} \right)_{\rm{micro}}$), and the subgrid meridional momentum fluxes due to vertical advection ($\left( v \right)^{\rm{subg-flux}}_{\rm{adv}}$).  
 The non-local parameterizations rely on different features for these two output variables.
 The prediction of $\left( {\partial q_p}/{\partial t} \right)_{\rm{micro}}$ relies locally on temperature and moisture, but non-locally on horizontal wind variables. Interestingly, for this output variable NN3D uses the non-local wind features to construct the horizontal wind divergence in order to approximate the vertical wind, and it barely relies on other non-local variables. 
 This result motivated us to train a single-column NN parameterization that includes the vertical wind as an input, and we find that such a single-column parameterization outperforms the non-local parameterization for $\left( {\partial q_p}/{\partial t} \right)_{\rm{micro}}$.
 By contrast, the prediction of  $\left( v \right)^{\rm{subg-flux}}_{\rm{adv}}$ relies both locally and non locally on the moisture and wind variables, and uses non-local wind information beyond that needed for the horizontal wind divergence. The non-local parameterization outperforms a single-column NN that includes the vertical velocity as an input for the output $\left( v \right)^{\rm{subg-flux}}_{\rm{adv}}$.
 Overall, we find that both non-local features and the local vertical velocity (or horizontal wind divergence) can substentially improve the offline performance of parameterizations, and that non-local features are especially important for the parameterization of subgrid momentum fluxes.

The large-scale horizontal wind and moisture divergence and the vertical velocity are both a cause and a consequence of convection, and there is debate as to whether these features should be included as an input to convection and cloud parameterizations \cite{emanuel1994large, george2021}. Detailed testing would be needed to determine how including these inputs in an ML parameterization affects simulation of the general circulation and transient disturbances, and whether their use in an ML parameterization may affect the robustness or numerical stability of the simulations.

In the context of developing parameterizations by coarse-graining output from high-resolution data, including the vertical velocity as an input is not equivalent to including the horizontal divergence as an input. The underlying reason is that the coarse-grained output does not necessarily exactly obey the same equations as the high-resolution data. As an example, one can consider the anelastic mass continuity equation, through which the coarse-grained vertical velocity is directly related to line averages of the horizontal winds on the boundaries of the grid cells, rather than to the coarse-grained horizontal winds. However, these line averages are effectively subgrid compared to the coarse-grained horizontal winds, and indeed, we find that a single-column NN parameterization that includes the vertical velocity as an input perform substantially better than a single-column NN parameterization that includes the horizontal wind divergence. This inconsistency between the coarse-grained output and the anelastic continuity equation, raises an interesting question as to which quantities can we expect to have similar statistics between a coarse-resolution simulation run with an ML parameterization and the coarse-grained output of the high-resolution simulation on which the parameterization was trained, even in the limit of a perfect parameterization.

Our offline results suggest that for some outputs the non-local parameterizations can be more accurate than a single-column parameterization at all latitudes, but in this work we mostly focused on characterizing cases with improvement at mid-latitudes, and further work is needed to explain the reasons for improvements in the tropics. Furthermore, we focused only on spatial non-locality and future studies should investigate the opportunity of using non-local information in time (e.g., \citeA{Han_2020_GCM}) which may be related to non-locality in the horizontal for propagating weather systems. Finally, our focus here is on investigating the potential of using non-local inputs (and the related issue of using the local divergence or vertical velocity as an input) and understanding the situations in which improved prediction occurs, and we leave to future work the important next step of implementing and testing such parameterization approaches in climate-model simulations.  We note that because the non-local inputs we used are in a close neighborhood, it is expected that they could be incorporated in parameterizations to improve simulations without greatly increasing computational expense. Another potential route forward is to adapt traditional (physics-based) parameterizations to take advantage of non-local inputs or horizontal gradients. 


\section*{Open Research}
\noindent Code and data used in this study are available at https://doi.org/10.5281/zenodo.6672908.


\acknowledgments
We thank Bill Boos for the output from the high-resolution simulation and Caroline Muller for the code to calculate CAPE consistent with SAM. 
This research received support by the generosity of Eric and Wendy Schmidt by recommendation of the Schmidt Futures as part of its its Virtual Earth System Research Institute (VESRI). 
PW acknowledges support from NSF 1906719. 
JY acknowledges support from the EAPS Houghton-Lorenz postdoctoral fellowship. 
PAO'G acknowledges support from NSF AGS 1749986. 


\bibliography{references}

\end{document}


%
%


\title{Supporting Information for ``Non-local parameterization of atmospheric subgrid processes with neural networks''}
%
%

%
%


\authors{Peidong Wang\affil{1}, Janni Yuval\affil{1}, and Paul A. O'Gorman\affil{1}}
\affiliation{1}{Department of Earth, Atmospheric and Planetary Sciences, Massachusetts Institute of Technology, Cambridge, \\Massachusetts 02139, USA}

\authorrunninghead{WANG, YUVAL, and O'GORMAN}
\titlerunninghead{NON-LOCAL PARAMETERIZATION}

%
%

%

\begin{article}

\noindent\textbf{Contents of this file}
\begin{enumerate}
\item Text S1 to S4
\item Figures S1 to S14

\end{enumerate}

\noindent\textbf{Introduction}

 Here we describe the layer-wise relevance propagation rules that we use (Text S1),  how we calculate the instantaneous precipitation rate (Text S2), how we calculate the convective available potential energy (Text S3), and how we calculate other measures for convective and symmetric instabilities (Text S4). Figure S1-S14 show supporting figures for the main text.

\clearpage

\noindent\textbf{Text S1. Layer-wise relevance propagation rules}

The general relevance propagation rule is as follows \cite{Montavon_LRP_review}:
 \begin{linenomath*}
 \begin{equation}
   R_j = \sum_{k}^{} \frac{a_j \cdot (w_{jk} + \gamma w_{jk}^{+}) }{\epsilon + \sum_{0,j^\prime}^{} a_j^\prime \cdot (w_{j^\prime k} + \gamma w_{j^\prime k}^{+})} R_k,
 \label{eqn_LRP_general}
 \end{equation}
 \end{linenomath*}
 where $R$ stands for the relevance, $j$ and $k$ are the neurons in two consecutive layers, with $j$ indexing a layer closer to the input layer and $k$ indexing a layer closer to the output layer, $a_j$ is the value propagated forward by neuron $j$ (activated by ReLU), $w_{jk}$ is the weight that connects neuron $j$ to $k$, and $w_{jk}^{+}$ is the same as $w_{jk}$ but only includes positive weights. The sum over $j^\prime$ is over all neurons in the layer indexed by $j$ but also including an extra neuron with index $0$ representing the bias. When back-propagating the relevance from the first hidden layer to the input layer, a different propagation rule is applied \cite{Montavon_LRP_input_layer}, and the relevance is calculated as:
 \begin{linenomath*}
 \begin{equation}
   R_j = \sum_{k}^{} \frac{w_{jk}^2}{\sum_{j^\prime}^{} w_{j^\prime k}^2} R_k,
 \label{eqn_LRP_input_layer}
 \end{equation}
 \end{linenomath*}
 where $j$ indexes the input in the input layer, and $k$ indexes the neuron in the first hidden layer.
There are two tunable parameters in LRP, namely, $\epsilon$ and $\gamma$, which reduce the noise and favors positive weights, respectively. 
We present the LRP results when both $\epsilon$ and $\gamma$ set to zero for simplicity,
but we verified that the results presented are not sensitive to different choices of the parameters $\epsilon$ and $\gamma$ (Figure~S11).

\noindent\textbf{Text S2. Calculation of instantaneous precipitation rate}

Following \citeA{Yani_2021GRL}, the instantaneous precipitation rate is estimated as:
 \begin{equation}
     P_{tot}(z=0)= \int_{0}^{\infty} \rho_0 \left( \frac{\partial q_p}{\partial t} \right)_{\rm{micro}} dz,
\label{precp_eqn}
 \end{equation}
 where $P_{\rm{tot}}(z=0)$ is the total precipitation rate at the surface (including all phases) and
$\left( {\partial q_p}/{\partial t} \right)_{\rm{micro}}$ is the tendency of the precipitating water mixing ratio.
We use this approach for the neural-network parameterization which does not make a separate prediction of the surface precipitation. 
We also use this approach to calculate the ``true'' instantaneous precipitation on the coarse grid by using the coarse-grained $\left( {\partial q_p}/{\partial t} \right)_{\rm{micro}}$ from the hi-res simulation. This was necessary because only 3-hourly averaged precipitation was stored as simulation output. 
In both cases, we only integrate over the bottom 30 vertical levels from 0 to 13.4 km which is the height range of the neural-network outputs.

\noindent\textbf{Text S3. Convective available potential energy (CAPE)}

Following \citeA{Muller_CAPE}, we calculate CAPE (with any convective inhibition removed) as:
\begin{equation}
    \textrm{CAPE} = \int_{0}^{z_{LZB}} g \left[ \frac{T_p-T_e}{T_e} + \left( \frac{R_v}{R_d} - 1 \right) \left( q_{v,p} - q_{v,e} \right) - \left( q_{c,p} - q_{c,e} \right) - \left( q_{i,p} - q_{i,e} \right) \right] dz, 
\label{CAPE_eqn}    
\end{equation}
where $z_{LZB}$ is the level of zero buoyancy, $g$ is the gravitational acceleration, $T_e$ is the temperature of the environment, $T_p$ is the temperature of the parcel following pseudo-adiabatic ascent, $R_v$ and $R_d$ are the gas constants for water vapor and dry air, respectively, $q$ is the mixing ratio where subscript $v$ stands for vapor, $c$ for cloud liquid, $i$ for ice, $e$ for environment, and $p$ for parcel.

\noindent\textbf{Text S4. Quantification of symmetric and convective instabilities}
 
 To provide measures of upright convective and symmetric instability, we first calculate the saturation moist static energy ($\textrm{MSE}^*$) as:
 \begin{equation}
 \textrm{MSE}^* = c_p T + g z + L q_v^*,
 \label{MSE_eqn}
 \end{equation}
 where $c_p$ is the specific heat of air at constant pressure, $T$ is the absolute temperature, $g$ is the gravitational acceleration, $z$ is the height, $q_v^*$ is the saturation vapor mixing ratio, and $L=L_c+L_f (1-\omega_n)$ with $L_c$ the latent heat of condensation, $L_f$ the latent heat of fusion, and $\omega_n$ the phase partition function for non-precipitating water in SAM.  $\textrm{MSE}^*$ is  conserved in SAM for saturated moist-adiabatic processes in the absence of melting and freezing of cloud condensates and precipitation.

 We diagnose conditional instability for upright convection at a given level when $ \partial \textrm{MSE}^*/\partial z < 0$. 
 Similarly, we diagnose conditional symmetric instability when the  saturation moist potential vorticity $\textrm{MPV}^*$ is negative \cite{Korty_2007}. We modify $\textrm{MPV}^*$ to use $\textrm{MSE}^*$ and the anelastic reference density profile as follows:
 \begin{equation}
 \textrm{MPV}^* = \frac{\textrm{sign}(f)}{c_p \rho_0(z)} \left( \nabla \times \mathbf{v} + f \hat{z}  \right) \cdot \nabla \textrm{MSE}^*,
 \label{MPV_eqn}
 \end{equation}
 where $\rho_0(z)$ is the reference density profile, $\mathbf{v}$ is the three-dimensional velocity, and $f$ is the Coriolis parameter on the equatorial $\beta$-plane. 
 By multiplying $\textrm{MPV}^*$ with the sign of $f$, the atmosphere is conditionally symmetrically unstable when $\mathrm{MPV}^* < 0$ in both hemispheres.  Bands of increased stability are evident near the melting lines in Figures S8 and S9, and this is related to melting and freezing of precipitation and condensate.

%
%
\end{article}
\clearpage

\begin{figure}
\label{train_test_seperation}
    \centering
    \includegraphics[width=1.0\linewidth]{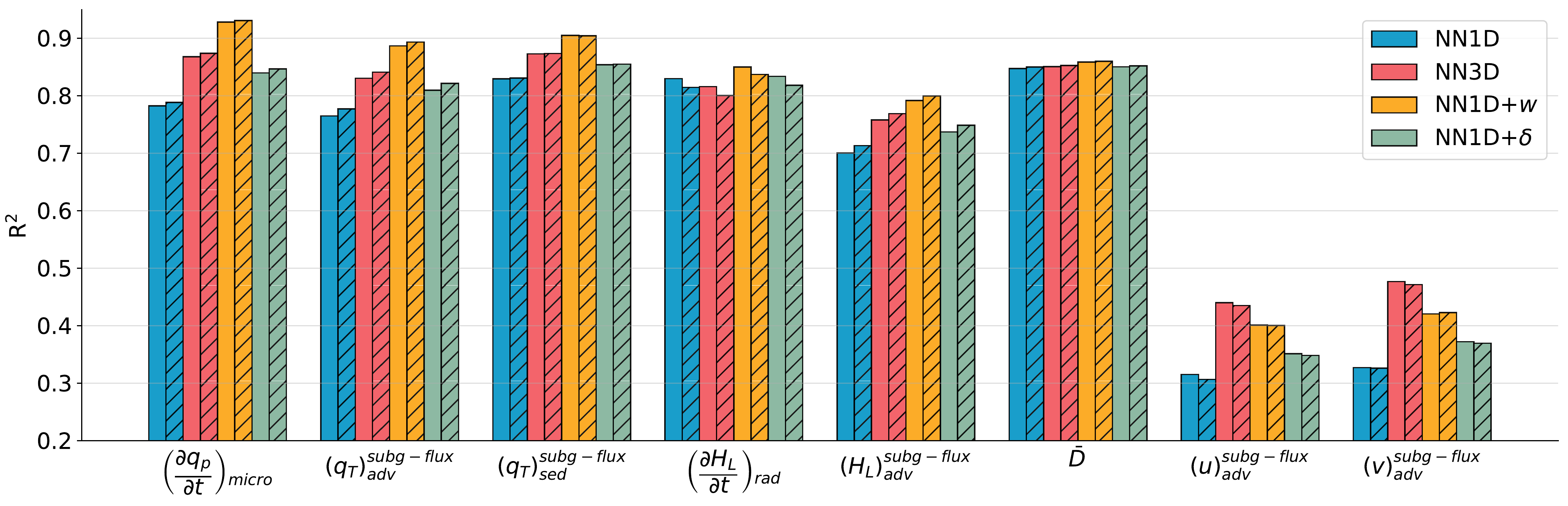}
    \caption{Similar to Figure 2, but also showing results for a different train/test split of the data. The bars without hatches are the same as the bars in Figure 2 (trained on 50\% of the data and tested on 40\% of the data), and the bars with hatches indicate more of the data is used for training (trained on 80\% of the data and tested on 10\% of the data).  The results show that NN performance is not hampered by training on a smaller portion of data.}
\end{figure}
\clearpage

\begin{figure}
\label{Loss_per_epoch}
\centering
\noindent\includegraphics[width=1.0\linewidth]{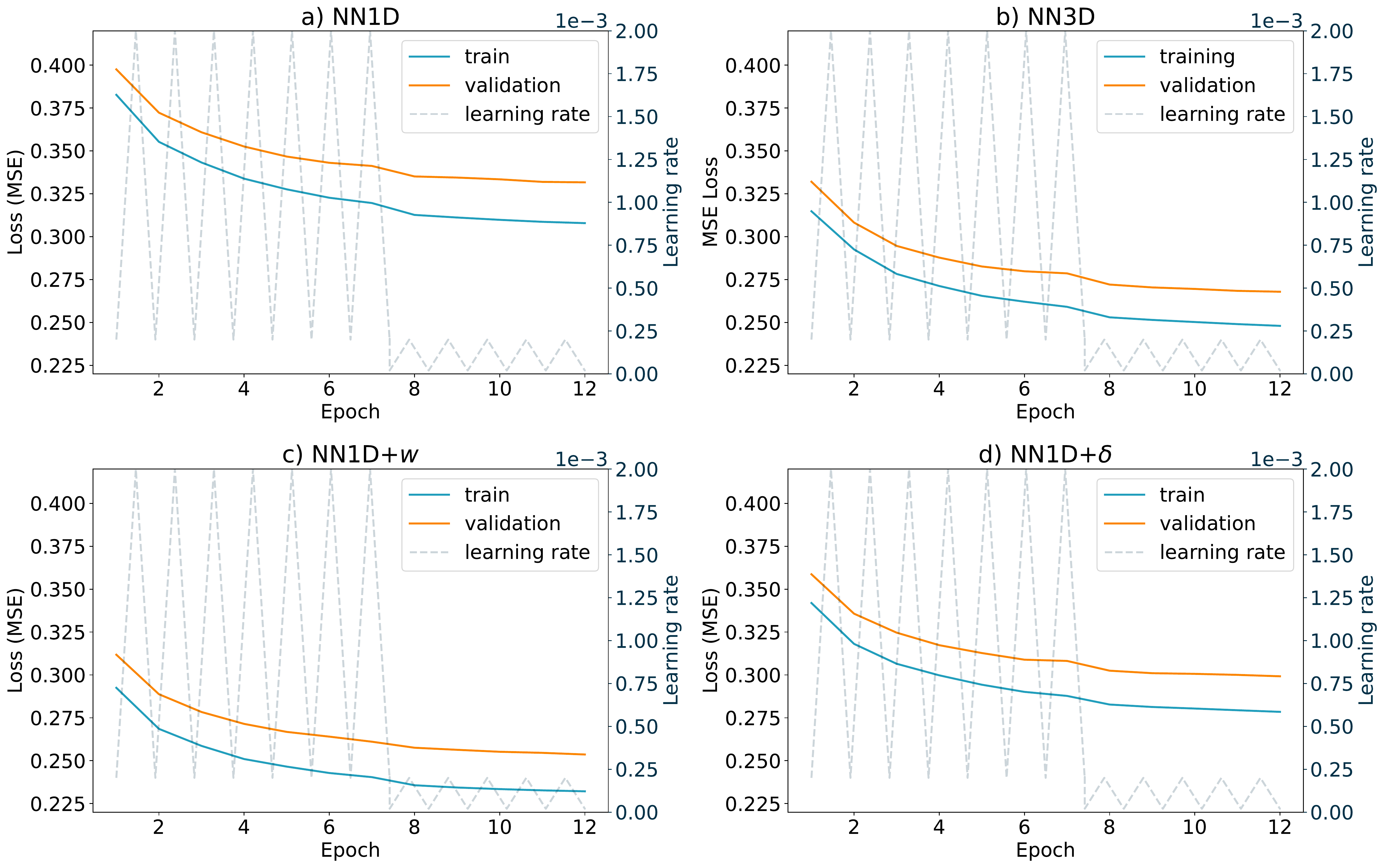}
\caption{Loss (mean-squared error) for the training  (blue) and validation (orange) sets at each epoch for the NNs used in Figure 2: \textbf{(a)} the single-column parameterization (NN1D), \textbf{(b)} the non-local parameterization (NN3D), \textbf{(c)} the single-column parameterization with local $w$ as an additional feature (NN1D+$w$), and \textbf{(d)} the single-column parameterization with local horizontal wind divergence $\delta$ as an additional feature (NN1D+$\delta$). 
Losses are plotted for the average of the batch at the end of each epoch. 
Learning rate is plotted in gray for each batch. 
The learning rate scheduler cycles through minimum and maximum learning rates once over each epoch. The maximum and minimum learning rates are reduced by a factor of 10 after training for 7 epochs. }
\end{figure}
\clearpage

\begin{figure}
\label{more_targets_remove_column_mean}
\centering
\noindent\includegraphics[width=1.0\linewidth]{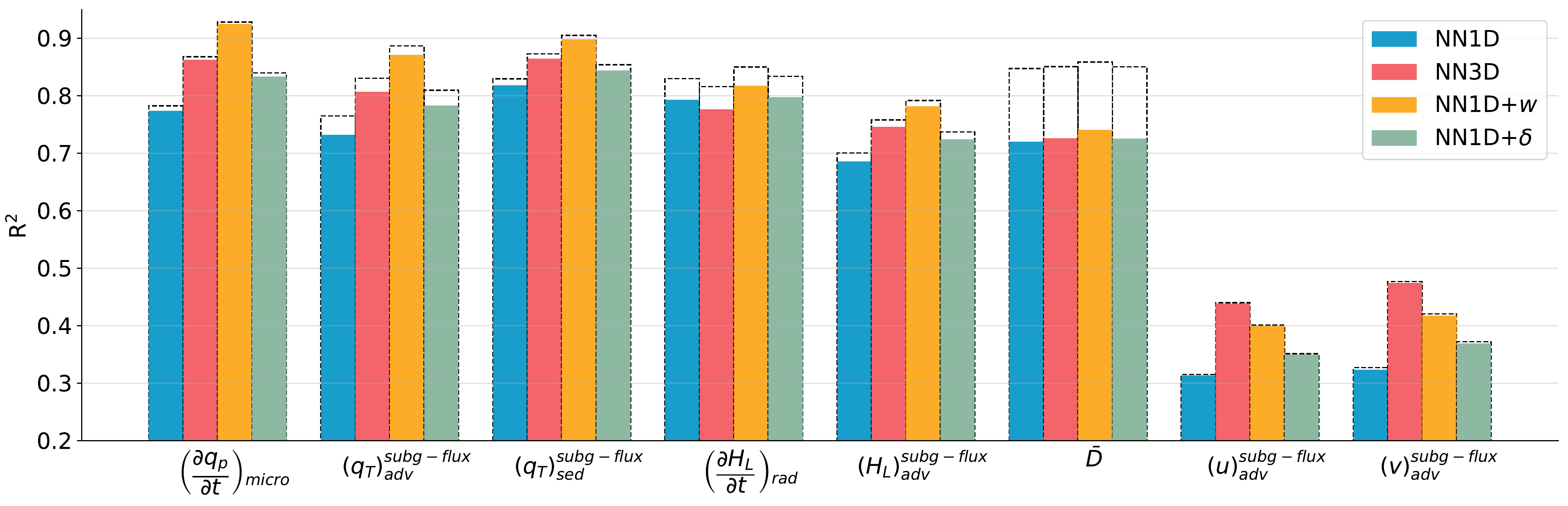}
\caption{Similar to Figure 2, but also showing R$^2$ values when the mean at each vertical level is first removed. The dashed bars are the same as the bars in Figure 2, and the colored bars indicate R$^2$ values when the mean at each vertical level is first removed.}
\end{figure}
\clearpage

\begin{figure}
\label{extreme_DQP}
\centering
\noindent\includegraphics[width=1.0\linewidth]{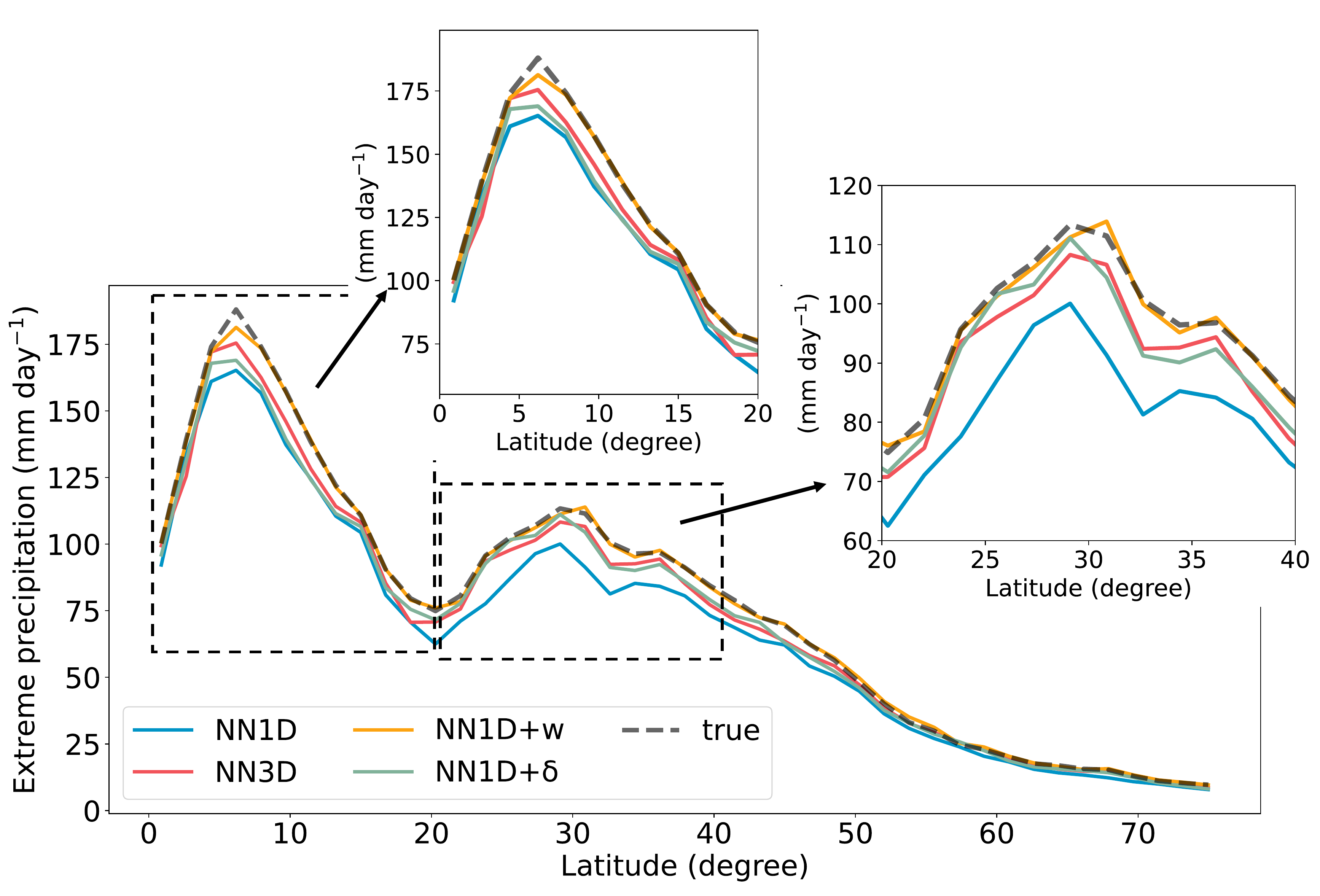}
\caption{The 99.9\textsuperscript{th} percentile of the instantaneous surface precipitation rate coarse-grained from the high-resolution simulation (gray dashed line), and from NNs with different input feature combinations: a single-column parameterization (NN1D; blue), a single-column parameterization with an additional 1D vertical velocity input (NN1D+$w$; yellow), a single-column parameterization with an additional 1D wind divergence input (NN1D+$\delta$; green), and a non-local parameterization (NN3D; red).
NN1D underestimates the extreme precipitation by up to 20\% at the mid-latitudes, while NN3D substantially improves the extreme precipitation in this region, and NN3D+$w$ improves the extreme precipitation at all latitudes.}
\end{figure}
\clearpage

\begin{figure}
\label{NN_compare_complexity}
\centering
\noindent\includegraphics[width=1.0\linewidth]{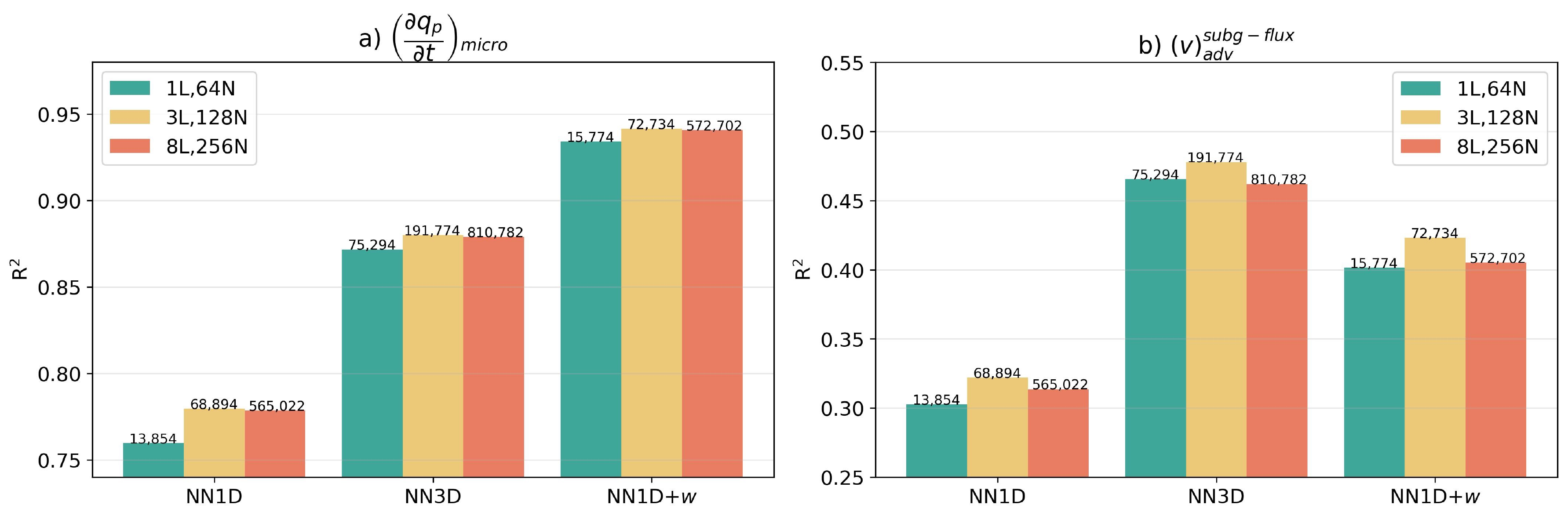}
\caption{The coefficient of determination (R$^2$) of the \textbf{(a)} tendency of the precipitating water mixing ratio due to microphysical processes ($\left( {\partial q_p}/{\partial t} \right)_{\rm{micro}}$) and \textbf{(b)} meridional momentum flux due to vertical advection ($\left( v \right)^{\rm{subg-flux}}_{\rm{adv}}$), from single-column parameterizations (NN1D), non-local parameterizations (NN3D) and single-column parameterizations with $w$ as additional input (NN1D+$w$). Different architectures are considered: 1 hidden layer and 64 neurons per layer (1L,64N); 3 hidden layers and 128 neurons per layer (3L,128N); 8 hidden layers and 256 neurons per layer (8L,256N). The number of trainable parameters associated with each neural network architecture is shown on top of each bar. 
NN3D and NN1D+$w$ with lower complexity (1L,64N) can significantly outperform NN1D with higher complexity (8L,256N), while using 87--97 \% fewer trainable parameters.}
\end{figure}
\clearpage


\begin{figure}
\label{QN_Vmom_midlat}
\centering
\noindent\includegraphics[width=1.0\linewidth]{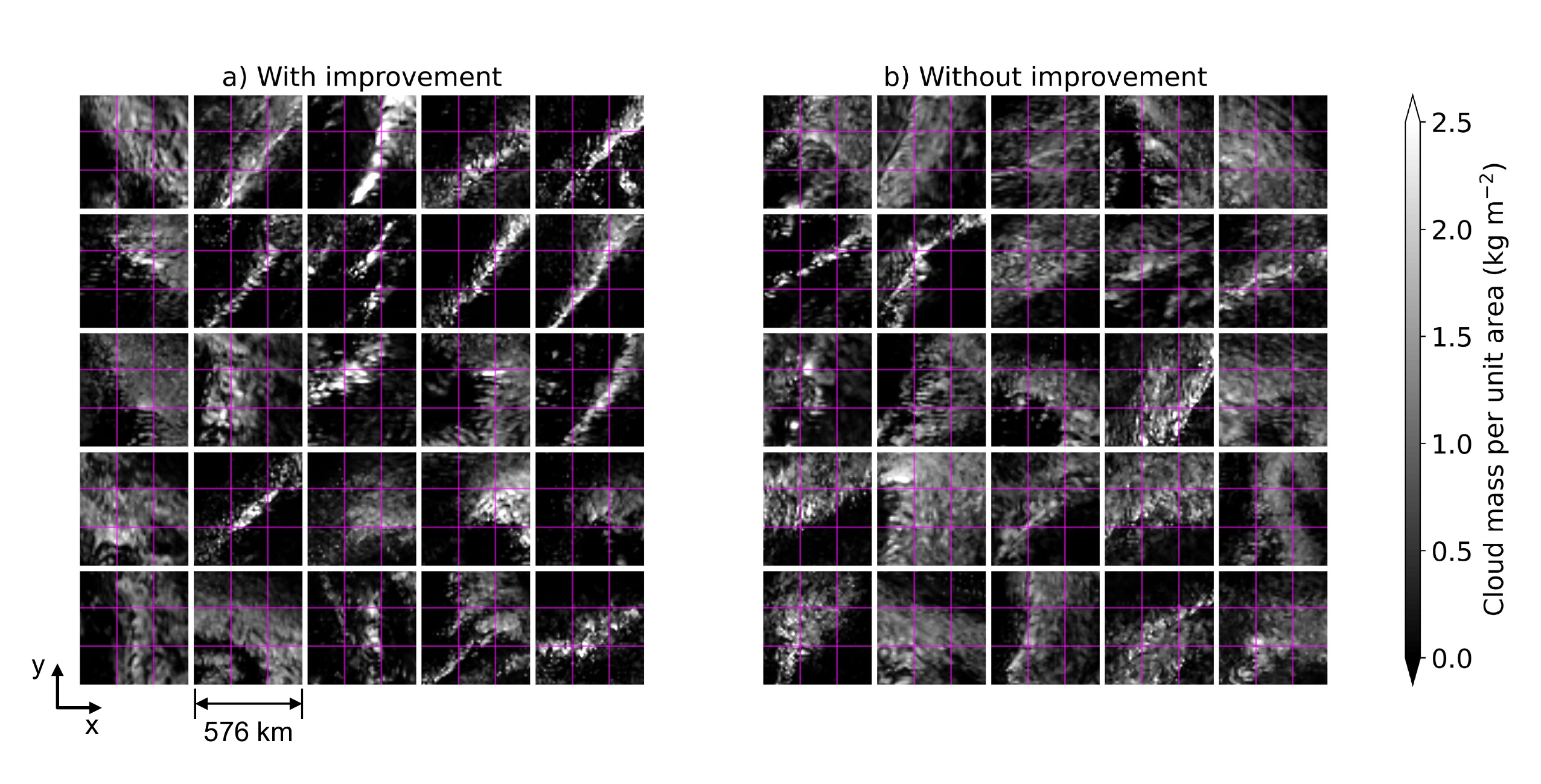}
\caption{Similar to Figure 4, but based on improvements in predicting $\left( v \right)^{\rm{subg-flux}}_{\rm{adv}}$.}
\end{figure}
\clearpage


\begin{figure}
\label{CAPE_Vmom_midlat}
\centering
\noindent\includegraphics[width=0.7\linewidth]{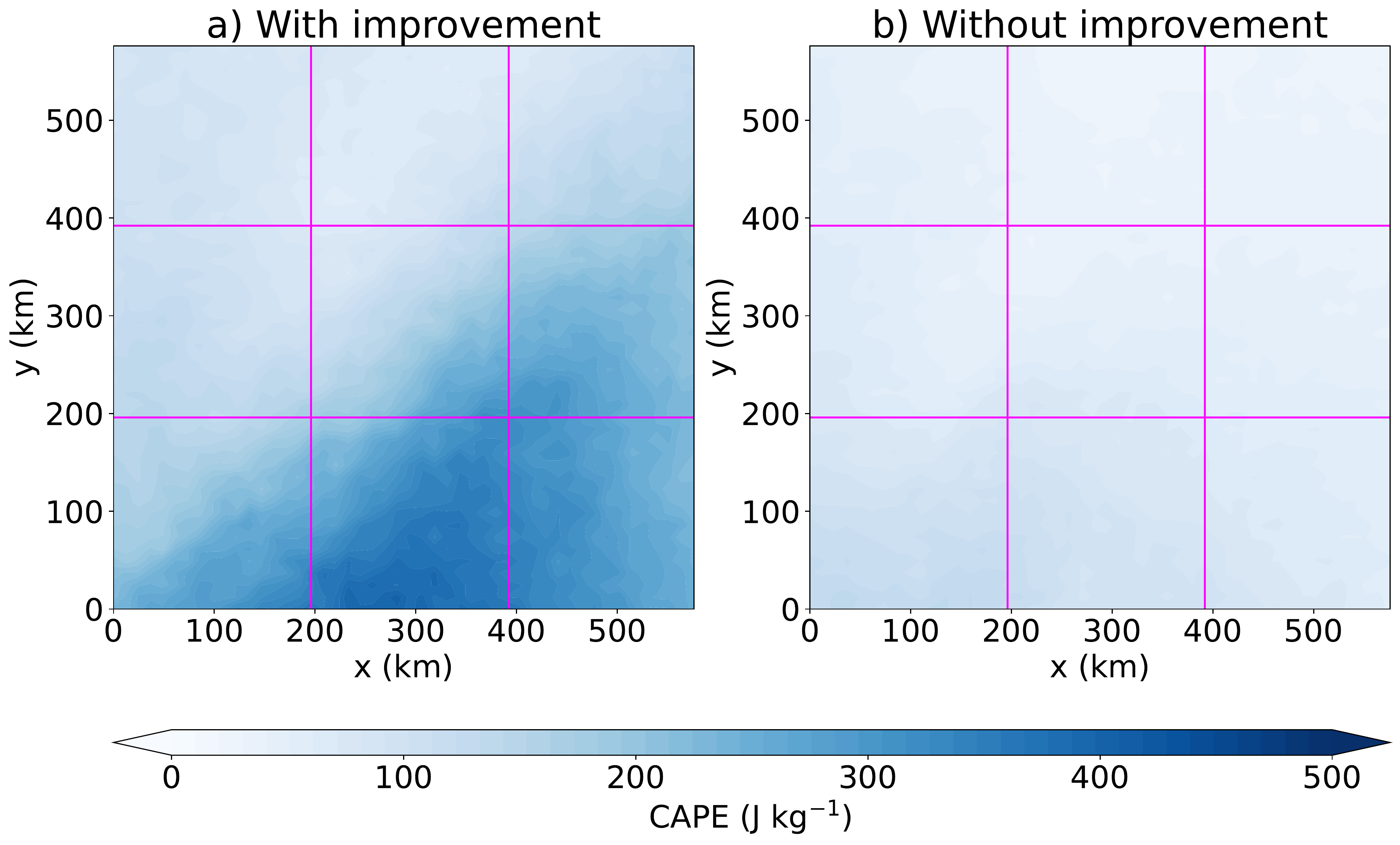}
\caption{Similar to Figure 5, but based on the improvement in predicting $\left( v \right)^{\rm{subg-flux}}_{\rm{adv}}$.}
\end{figure}
\clearpage

\begin{figure}
\label{midlat_MPV_DQP}
\centering
\noindent\includegraphics[width=1.0\linewidth]{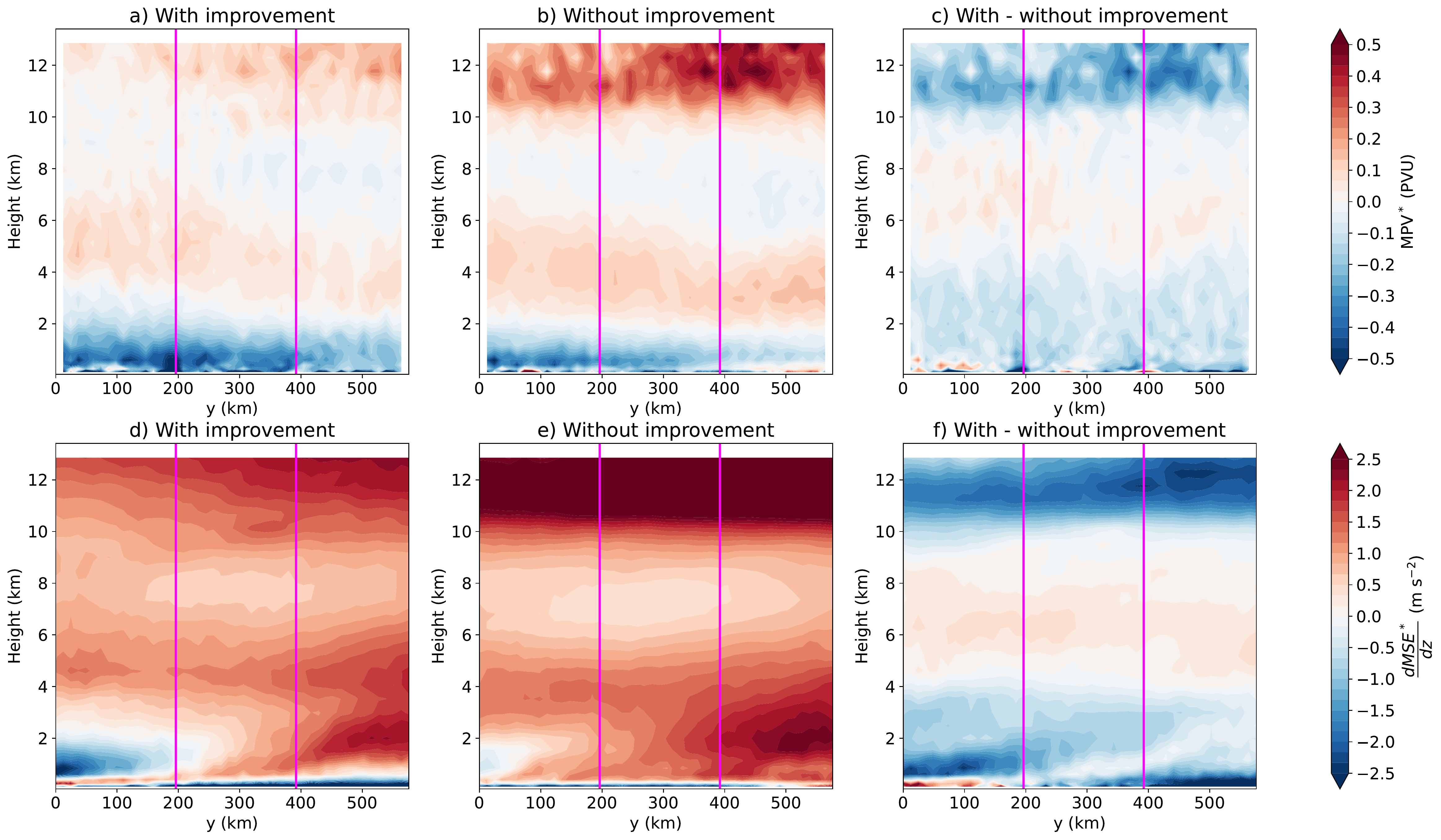}
\caption{Different measures of instability based on the cases with largest/little improvement when predicting $\left( {\partial q_p}/{\partial t} \right)_{\rm{micro}}$ using non-local parameterization compared to local parameterization. \textbf{(a-c)} The saturation potential vorticity (which is a measure of conditional symmetric instability) and \textbf{(d-f)} the vertical derivative of the moist static energy (which is a measure of conditional instability to upright convection) as a function of height and meridional distance. These measures of instability are calculated from the high resolution data for the 300 mid-latitude cases that \textbf{(a,d)} have the largest improvements when the non-local parameterization (NN3D) is used, \textbf{(b,e)} have little improvement when NN3D is used, and \textbf{(c,f)} for the difference between the cases with largest and little improvement. See Section 3.2 for details of how the groups of cases are chosen. Averages are taken in longitude over three coarse columns and over the cases in question. Magenta lines show the edges of the coarse columns.  An increased band of static stability is evident near 4 -- 5 km due to the melting level. Both instability measures suggest that the atmosphere is overall less stable in cases with largest improvements, especially below 4 km.}
\end{figure}
\clearpage


\begin{figure}
\label{midlat_MPV_Umom}
\centering
\noindent\includegraphics[width=1.0\linewidth]{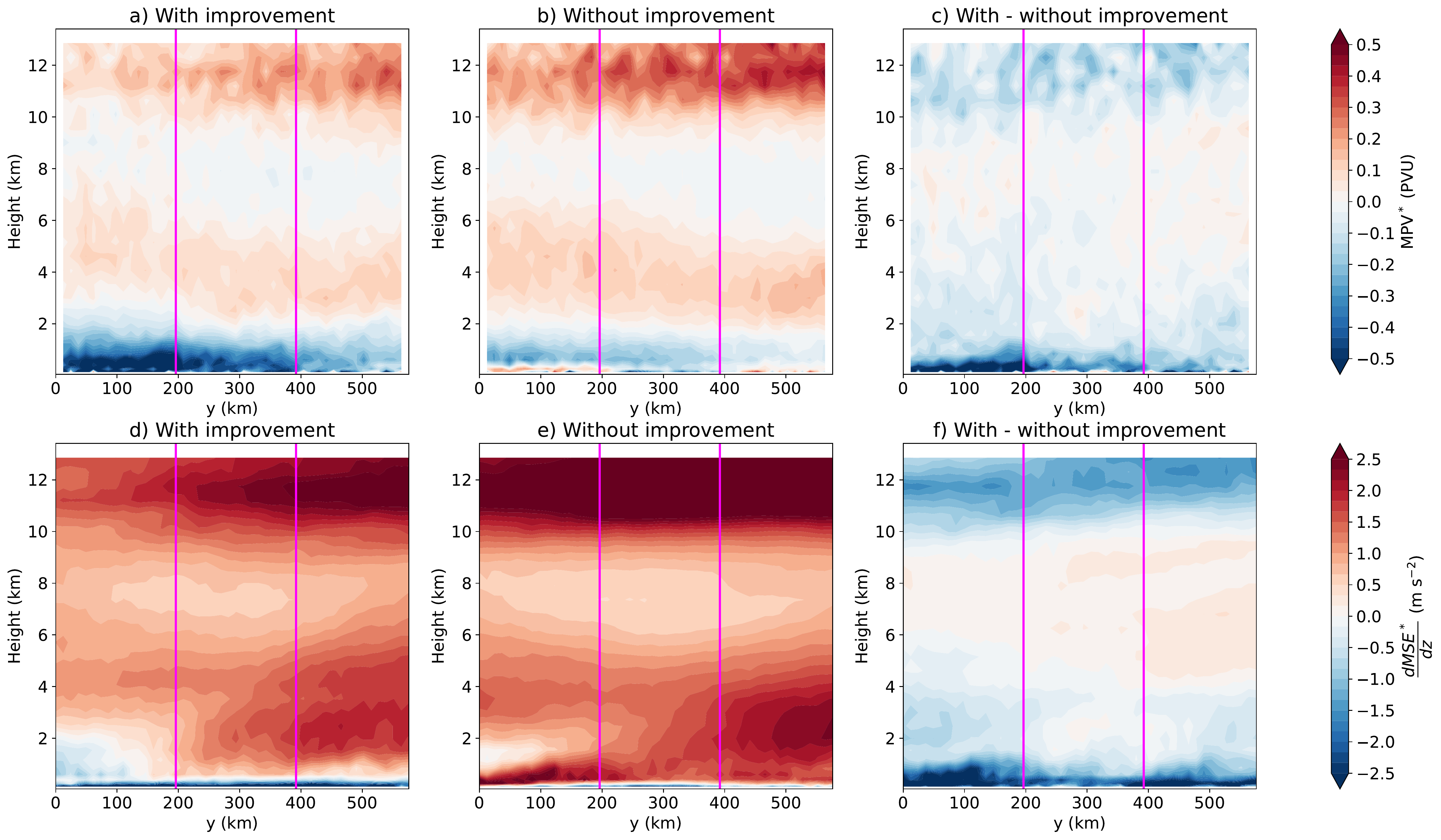}
\caption{Similar to Figure S8, but based on the improvement in predicting $\left( v \right)^{\rm{subg-flux}}_{\rm{adv}}$.}
\end{figure}
\clearpage

\begin{figure}
 \centering
 \includegraphics[width=0.85\linewidth]{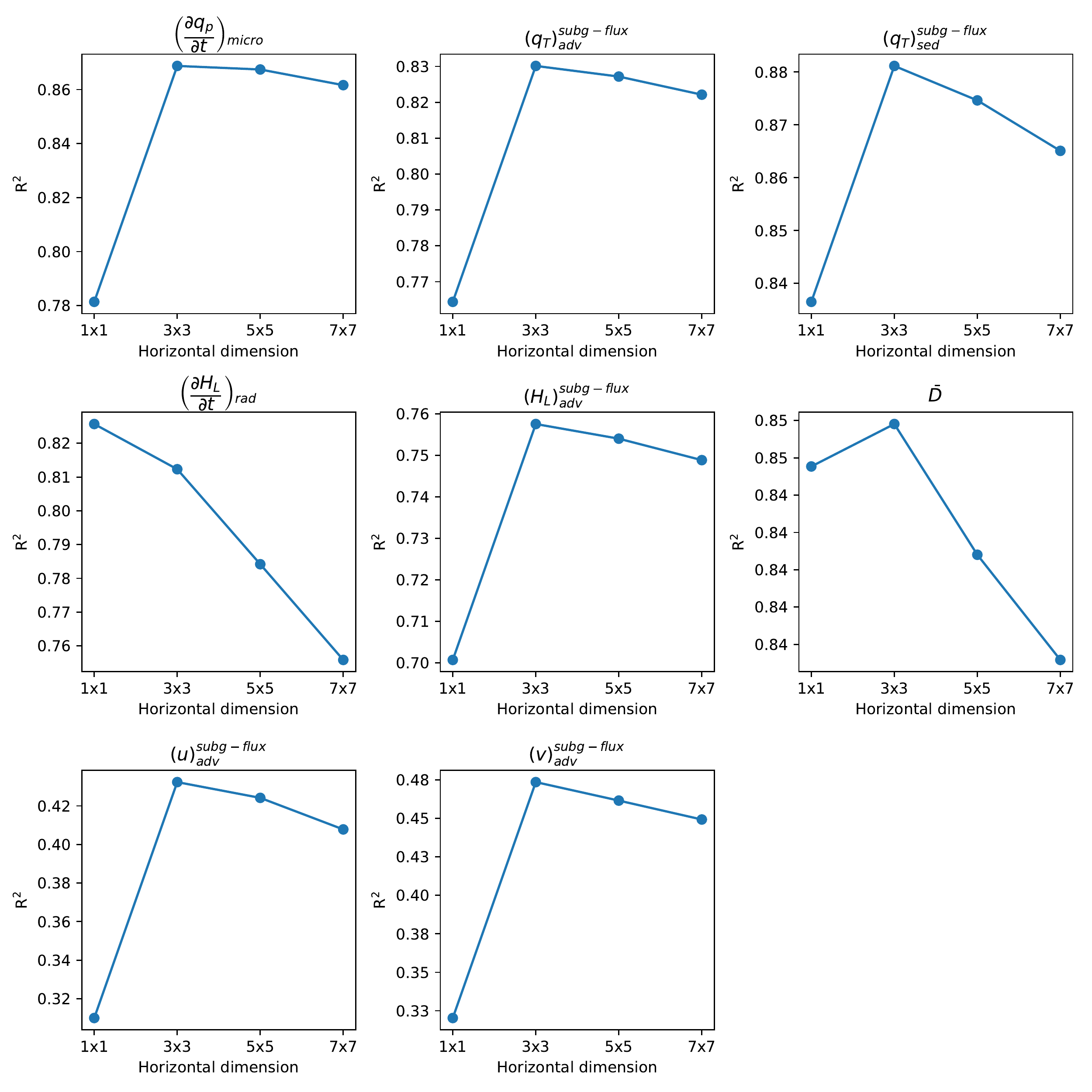}
 \caption{Global R$^2$ values  for NNs  using input features of $T, q_T, u, v$ versus the number of non-local grid boxes used as inputs, ranging from 1$\times$1 (local parameterization) up to 7$\times$7 non-local grid boxes. To be consistent in the number of training/testing samples between NNs with different level of non-locality, we exclude the three outermost columns of data in each snapshot. Therefore, each snapshot has 2,520 samples instead of 2,752 samples as was used for results in other figures.}
 \label{fig_diff_coarse_factor}
 \end{figure}
 \clearpage

\begin{figure}
\label{midlat_LRP_hyperparam}
\centering
\noindent\includegraphics[width=1.0\linewidth]{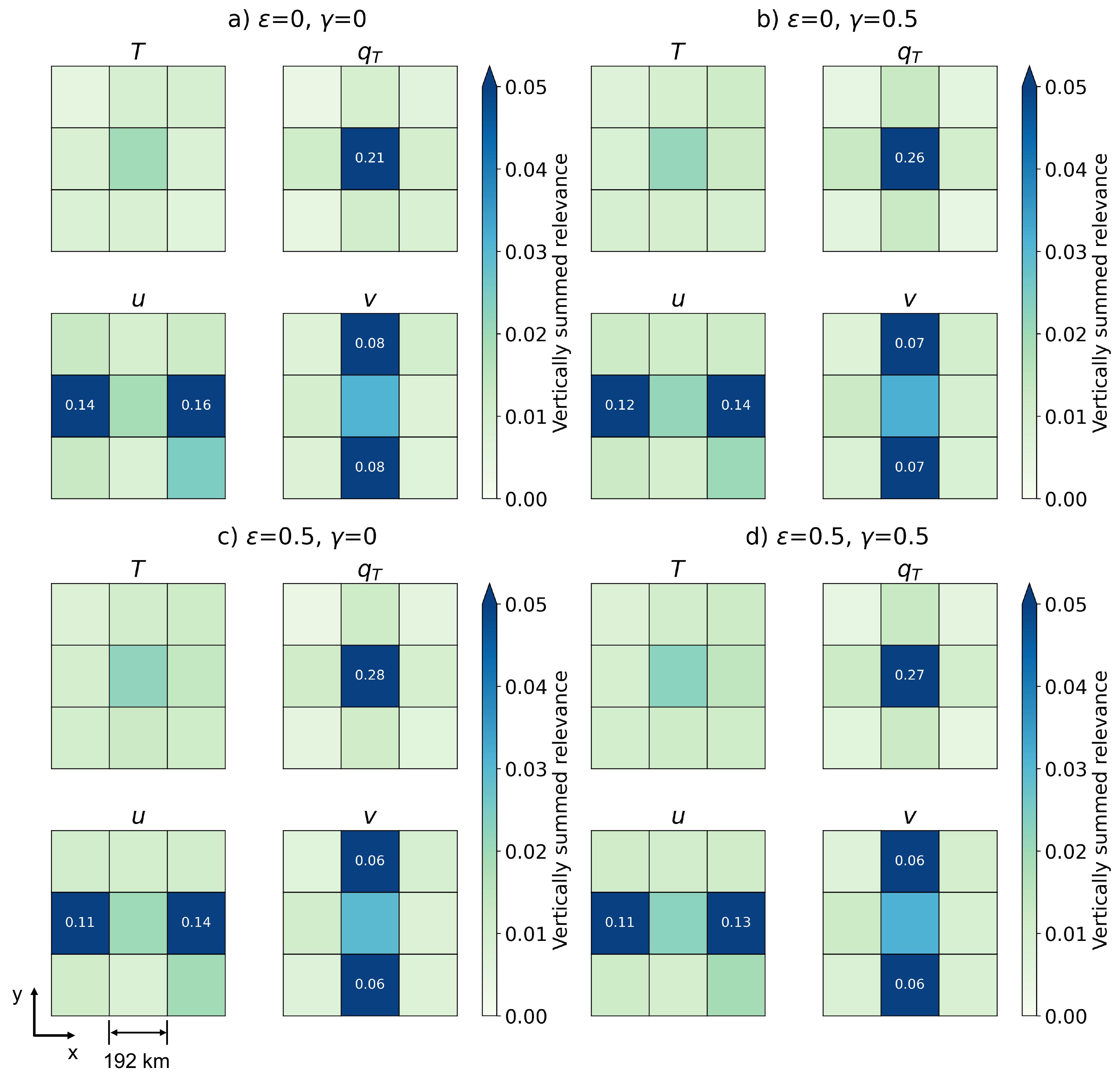}
\caption{Similar to Figure 6a, but for different $\epsilon$ and $\gamma$ values used to calculate relevance. 
Equation~\ref{eqn_LRP_general} in the supplement shows how $\epsilon$ and $\gamma$ enter into the calculation of the relevance. 
Larger $\epsilon$ values tend to reduce noise in relevance, and larger $\gamma$ values favor positive contributions from neurons. This figure shows that the exact relevance score can be slightly different using various parameters, but the overall pattern remains the same.}
\end{figure}
\clearpage

\begin{figure}
\label{QN_postage_tropics_DQP}
\centering
\noindent\includegraphics[width=1.0\linewidth]{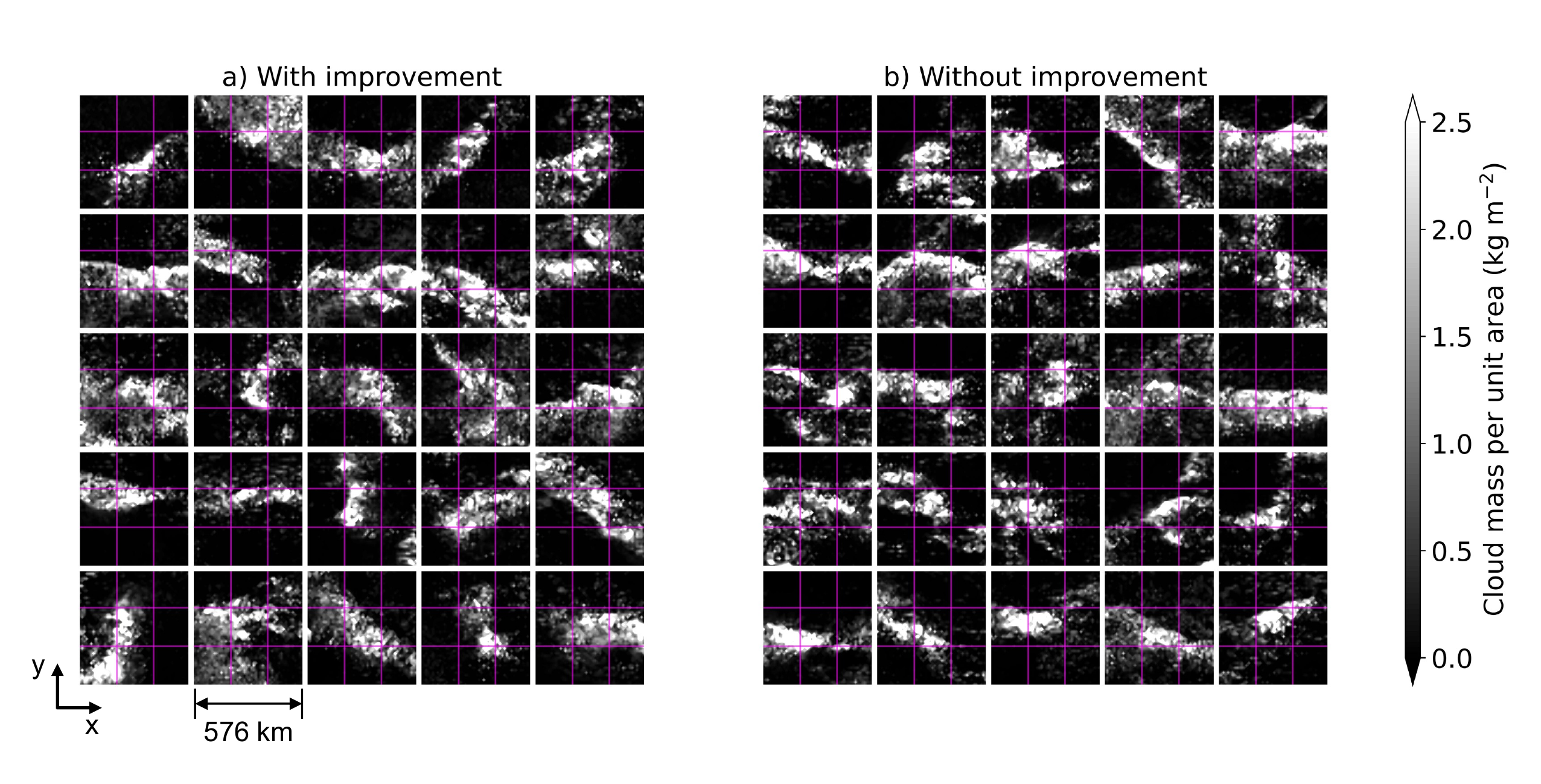}
\caption{Similar to Figure 4 but for the tropical cases when predicting $\left( {\partial q_p}/{\partial t} \right)_{\rm{micro}}$. Unlike the mid-latitude cases, cloud shapes in the cases with and without improvement do not obviously differ. }
\end{figure}
\clearpage



\begin{figure}
\label{CAPE_tropics_DQP}
\centering
\noindent\includegraphics[width=0.7\linewidth]{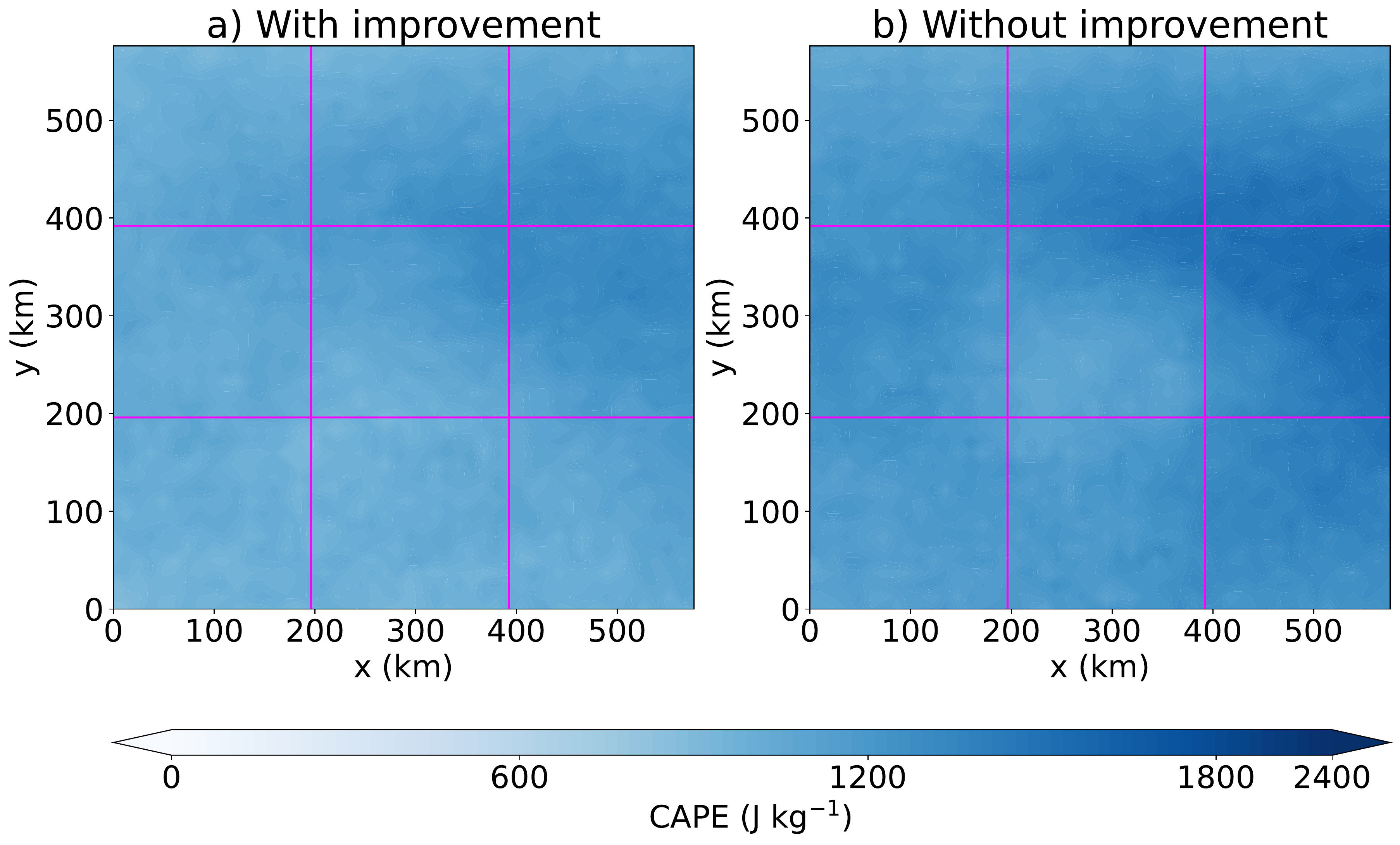}
\caption{Similar to Figure 5, but for tropical cases when predicting $\left( {\partial q_p}/{\partial t} \right)_{\rm{micro}}$. Cases with improvement have slightly lower CAPE than cases without improvement, suggesting the atmosphere is more stable to convection, which is opposite compared to the mid-latitudes.}
\end{figure}
\clearpage

\begin{figure}
\label{tropics_LRP}
\centering
\noindent\includegraphics[width=1\linewidth]{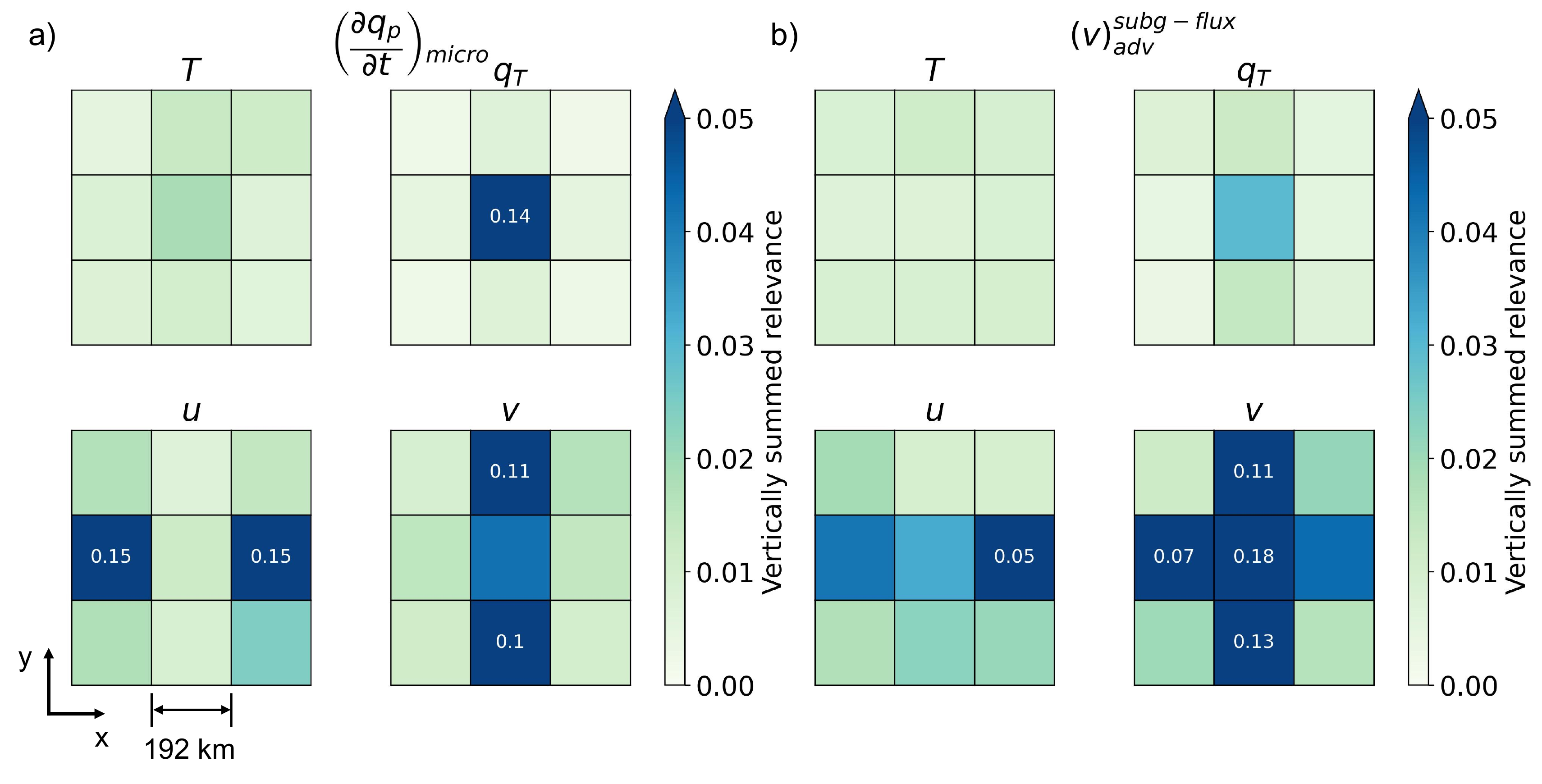}
\caption{Similar to Figure 6, but for tropical cases. The exact relevance values are different, but the overall  patterns are similar, suggesting that similar information is being used.}
\end{figure}
\clearpage



\bibliography{references}